\documentclass[fleqn,10pt]{olplainarticle}

\graphicspath{ {images/} }
\usepackage{newfloat}
\DeclareFloatingEnvironment[name={Supplementary Figure}]{suppfigure}
\DeclareFloatingEnvironment[name={Supplementary Table}]{supptable}
\usepackage{url} 
\makeatletter  
\renewcommand\listoffigures{%
        \@starttoc{lof}%
}
\makeatother

\title{A new approach for extracting information from protein dynamics}

\author[1]{Jenny Liu}
\author[2,*]{Lu\'{i}s A. N. Amaral}
\author[1,*]{Sinan Keten}
\affil[1]{Department of Mechanical Engineering, Northwestern University}
\affil[2]{Department of Chemical and Biological Engineering, Northwestern University}
\affil[*]{These authors contributed equally.}

\begin{document}

\flushbottom
\maketitle
\thispagestyle{empty}

\section*{Network inference from protein dynamics}

\section*{Abstract}
Increased ability to predict protein structures is moving research focus towards understanding protein dynamics. A promising approach is to represent protein dynamics through networks and take advantage of well-developed methods from network science. Most studies build protein dynamics networks from correlation measures, an approach that only works under very specific conditions, instead of the more robust inverse approach. Thus, we apply the inverse approach to the dynamics of protein dihedral angles, a system of internal coordinates, to avoid structural alignment. Using the well-characterized adhesion protein, FimH, we show that our method identifies networks that are physically interpretable, robust, and relevant to the allosteric pathway sites. We further use our approach to detect dynamical differences, despite structural similarity, for Siglec-8 in the immune system, and the SARS-CoV-2 spike protein. Our study demonstrates that using the inverse approach to extract a network from protein dynamics yields important biophysical insights.

\section*{Keywords}
Molecular Dynamics Simulation; Protein; SARS-CoV-2; Sialic Acid Binding Immunoglobulin-like Lectins; Fimbrial Adhesins

\section*{Statements}
Data available on request from the authors. We have no conflict of interest to declare.

\section*{Acknowledgments}
The authors thank Martin Gerlach and Kerim Dansuk for helpful conversations. J.L. thanks the Paul and Daisy Soros Fellowship, the Northwestern Quest High Performance Computing Cluster, and the National Institute of Health T32GM008152. This project was also supported by the Office of Naval Research N00014163175 and N000141512701 (S.K.), the National Science Foundation 2034584 (S.K.) and 1764421-01 (L.A.N.L.), and the Simons Foundation 597491-01 (L.A.N.L.).

\clearpage
\section*{Introduction}

Advances in experimental structure determination \citep{Levitt2007GrowthData, Terwilliger2009LessonsGenomics},
computational structure prediction \citep{Jumper2021HighlyAlphaFold},
and molecular dynamics simulations \citep{Hollingsworth2018MolecularAll}
have set the stage for high-throughput characterization of protein dynamics using molecular dynamics (MD) simulations.
Combined with increased computational power, these advances have led to rapidly increasing numbers of longer MD simulations for larger macromolecular systems \citep{Lee2018ExascaleBiology}. As a result, large datasets of MD trajectories are available from individual research labs \citep{Lee2018ExascaleBiology} and repositories such as MoDEL \citep{Meyer2010MoDELTrajectories},
Dynameomics \citep{Rysavy2014Dynameomics:Prediction}, Dryad, NoMaD, and MolSSI \citep{Elofsson2019TenSystems}.

This wealth of MD trajectory data creates opportunities for expanding our understanding of protein dynamics and function. While snapshots from MD trajectories contain information about low energy states and can be used to identify conformational changes, some physical phenomena -- such as dynamic allostery in proteins -- may be better characterized by dynamics over the course of the trajectory \citep{Wodak2019AllosteryApplications}. MD simulations can capture differences among protein variants, the impact of mutations, and modulation by small molecule binding at spatiotemporal resolutions that are difficult, or even impossible, to obtain experimentally \citep{Hollingsworth2018MolecularAll}.

Taking advantage of this growing wealth of MD trajectory data will require the development of robust methods for automated analysis. A common strategy for analyzing dynamics data involves creating a network by directly calculating contact times and interaction energies \citep{Amor2016PredictionPropensities, DiPaola2013ProteinChemistry, Yao2019EstablishingAnalysis, Sercinoglu2018GRINN:Simulations}. An alternative strategy used in network science aims to identify the underlying interactions of multi-component systems by inferring a network structure from dynamics \citep{Nguyen2017InverseScience}. Compared to interaction energy networks, building networks from dynamics is typically less computationally expensive and avoids less rigorous modeling of water or entropic contributions to the free energy \citep{Wang2019End-PointDesign}. Identifying a network structure makes it possible to apply network analysis tools in order to uncover emergent properties such as densely-connected communities \citep{DiNola1984FreeDynamics}, hotspots with many edges \citep{Singh2017QuantifyingCARDS}, and paths connecting active sites and allosteric regulatory sites in distant protein regions \citep{DiPaola2013ProteinChemistry, Wang2018InferringNetworks}.

In the study of proteins, the typical approach for constructing networks from protein MD simulations has made use of correlation measures that quantify how different protein regions "move together." These include a variety of methods that use linear \citep{Bowerman2016DetectingSimulation} and non-linear \citep{Hacisuleyman2017EntropyUbiquitin, Gasper2012AllostericActivities, Melo2020GeneralizedTrajectories, Lange2006GeneralizedDynamics, DuBay2011Long-rangeAlone, Singh2017QuantifyingCARDS} correlation measures. Yet a rigorous mathematical analysis demonstrates that inferring the network structure by solving the inverse problem for a system that could produce the observed correlations is a more accurate approach than using the correlations directly \citep{Nguyen2017InverseScience}. The most straightforward form of solving the inverse problem is simply calculating the inverse of the covariance matrix \citep{Nguyen2017InverseScience}. Here we apply the inverse covariance approach to MD trajectory data. 

A remaining challenge is how to define the nodes in such a network representation. An approach that has been used in proteins is to assign a node to each C$\alpha$ atom in elastic network models \citep{Atilgan2001AnisotropyModel, Moritsugu2007Coarse-grainedHessian} . This choice has some appeal because the model describes ``beads", located in Cartesian coordinates, connected by linear springs \citep{Atilgan2001AnisotropyModel, Moritsugu2007Coarse-grainedHessian}. However, using Cartesian coordinates requires a structural alignment step that can introduce "artifacts" during hinged motion for multi-domain proteins, and even for small, single-domain proteins \citep{Altis2008ConstructionAnalysis}. Previously, we have demonstrated that an internal coordinate system using dihedral angles makes it possible to accurately localize motion that affects elements distal to the hinge in fimH \citep{Liu2021ConformationalMutation}.

Here, we show that network inference using inverse covariance analysis is robust across replicates and that it uncovers strong interactions among backbone dihedrals that form a contact-map pattern. While the contact-map pattern is also seen in elastic network models for single domains with high conformational stability, we continue to see this pattern even for multi-domain proteins with hinged motion when using inverse covariance analysis.

We demonstrate the value of the proposed approach by studying three physiologically significant proteins: the bacterial adhesion protein, FimH\textsubscript{L}, the human immune adhesion protein Siglec-8 \citep{Propster2016StructuralSiglec-8},
and two domains of the SARS-CoV-2 spike protein involved in adhesion to the human ACE2 receptor \citep{Wrapp2020Cryo-EMConformation}.
In addition to comparing the different structures of wild type and mutant FimH\textsubscript{L}, we are also able to detect localized structural changes due to breaking a disulfide bond \textit{in silico}.
For Siglec-8, we are able to detect differences between ``apo'' and ``holo'' states, despite their structural similarity \citep{Propster2016StructuralSiglec-8}.
For the SARS-CoV-2 spike protein, we examined the receptor binding domain (RBD) and its connecting subdomain 1 (SD1). While the hinge region connecting RBD-SD1 is open in the up state and closed in the down state, the individual domains remain structurally similar in the up and down states \citep{Wrapp2020Cryo-EMConformation}.
For Siglec-8 and spike RBD-SD1, which do not have large structural changes within protein domains, comparing inferred networks allowed us to identify dynamical changes and contributions to stability. 

\section*{Materials and Methods}

\subsection*{Protein structures}
We retrieved crystal structure for FimH\textsubscript{L} wild type and mutant, Siglec-8 apo and holo, and SARS-CoV-2 spike protein from the Protein Data Bank, as detailed in Table \ref{tab:pdb}. For FimH\textsubscript{L}, we used crystal structures of the lectin domain without ligand for both the wild type and the mutant. To compare dynamics with and without the disulfide bond as a local perturbation, we used Visual Molecular Dynamics (VMD) to define the bond or two cystines for FimH\textsubscript{L}. For Siglec-8, we simulated the ligand 6'S sLe\textsuperscript{x} without the 3-amino-propyl linker, which is not thought to interact with the binding pocket \citep{Propster2016StructuralSiglec-8}.

For the SARS-CoV-2 spike protein, we started from the refined structure on the CHARMM-GUI archive \citep{Woo2020DevelopingMembrane, Wrapp2020Cryo-EMConformation}. To focus on one hinge system that is thought to be different between the down and up states, we isolated the receptor binding domain (RBD) and subdomain 1 (SD1) protein subunits without the glycans. For the up state, we used chain A where the RBD is accessible for binding the ACE2 receptor on human cells \citep{Wrapp2020Cryo-EMConformation}, and for the down state, we used chain B. We used the structure

We prepared all systems using VMD version 1.9.3 \citep{Humphrey1996VMD:Dynamics}. We solvated each protein with at least 16 \AA\ of TIP3P water molecules on each side to prevent interactions with itself through the periodic boundary conditions. We added sodium and chloride ions to neutralize the system and achieve the desired salt concentration in Table \ref{tab:pdb}.

\subsection*{MD simulations}
We performed all-atomistic MD simulations using NAMD \citep{Phillips2005ScalableNAMD}, with the CHARMM force field\citep{Huang2017CHARMM36m:Proteins}. Our NAMD simulation parameters and system details are listed in Table \ref{tab:namd}.

\begin{table}[ht]
\caption{\textbf{PDB structures studied (RRID:SCR\_012820).} We studied FimH\textsubscript{L} in the active (wild type) and inactive (Arg60Pro mutant) states; the human immune-inhibitory protein Siglec-8 in the apo and holo (6'S sLe\textsuperscript{x}-bound) states; and the SARS-CoV-2 spike protein RBD-SD1 domains in the up and down states.}
\label{tab:pdb}
\begin{tabular}{rcccccccc}
\textbf{Protein}                      & \multicolumn{3}{c}{\textbf{FimH}}                                                & \multicolumn{2}{c}{\textbf{Siglec-8}}             & \multicolumn{3}{c}{\textbf{Spike RBD-SD1}} \\ \cline{1-9} 
\textbf{Type}                         & \multicolumn{3}{c|}{Crystal}                                                     & \multicolumn{2}{c|}{NMR}                          & \multicolumn{3}{c}{Cryo-EM}                \\
\textbf{{[}NaCl{]} (mM)}              & \multicolumn{3}{c|}{50}                                                          & \multicolumn{2}{c|}{150}                          & \multicolumn{3}{c}{150}                    \\
\textbf{State}                        & \textbf{Wild type} & \textbf{Mutant} & \multicolumn{1}{c|}{\textbf{Full-length}} & \textbf{Apo} & \multicolumn{1}{c|}{\textbf{Holo}} & \textbf{Up} & \textbf{Down} & \textbf{Off} \\
\textbf{PDBID}                        & 4AUU               & 5MCA            & \multicolumn{1}{c|}{4XOD}                 & 2N7A         & \multicolumn{1}{c|}{2N7B}          & 6VSB        & 6VSB          & 6VXX         \\
\textbf{Protein residues}             & 158                & 158             & \multicolumn{1}{c|}{279}                  & \multicolumn{2}{c|}{145}                          & \multicolumn{3}{c}{276}                    \\
\textbf{Protein atoms}                & 2,360               & 2,350            & \multicolumn{1}{c|}{4,270}                 & \multicolumn{2}{c|}{2,290}                         & \multicolumn{3}{c}{4,286}                   \\
\textbf{System atoms}                 & 32,292              & 31,917           & \multicolumn{1}{c|}{60,376}                & 42,684        & \multicolumn{1}{c|}{50,879}         & 80,383       & 89,236         & 78,740        \\
\multirow{3}{*}{\textbf{System size}} & 94                 & 87              & \multicolumn{1}{c|}{123}                  & 97           & \multicolumn{1}{c|}{111}           & 125         & 116           & 115          \\
                                      & 61                 & 67              & \multicolumn{1}{c|}{72}                   & 70           & \multicolumn{1}{c|}{72}            & 94          & 95            & 94           \\
                                      & 59                 & 59              & \multicolumn{1}{c|}{72}                   & 67           & \multicolumn{1}{c|}{67}            & 81          & 81            & 77          
\end{tabular}
\end{table}

\begin{table}[ht]
\caption{\textbf{Details of molecular dynamics simulations.}}
\label{tab:namd}
\begin{tabular}{ll}
\textbf{Parameter}                  & \textbf{Value}                                                                                                                                                \\
\textbf{Setup}                      & VMD 1.9.3 (RRID:SCR\_004905)                                                                                                                                   \\
\textbf{Simulation engine}          & NAMD 2.13 (RRID:SCR\_014894)                                                                                                                                    \\
\textbf{Ensemble}                   & NPT                                                                                                                                                           \\
\textbf{Temperature}                & 300 K                                                                                                                                                         \\
\textbf{Pressure}                   & 1 atm                                                                                                                                                         \\
\textbf{Non-bonded interactions}    & Lennard-Jones potential ( cutoff)                                                                                                                             \\
\textbf{Electrostatic interactions} & Particle-Mesh Ewald sum method                                                                                                                                \\
\textbf{Forcefield}                 & CHARMM c36 July 2018 update                                                                                                                                   \\
\textbf{Timestep}                   & 1 fs                                                                                                                                                          \\
\textbf{Coordinate saved every}     & 1 ps                                                                                                                                                          \\
\textbf{Energy minimization}        & \begin{tabular}[c]{@{}l@{}}Conjugate gradient algorithm in NAMD\\ $\geq$ 10,000 steps with protein fixed\\ $\geq$ 10,000 steps with protein free\end{tabular}
\end{tabular}
\end{table}

After observing differences in correlated protein motions between replicates, we performed three replicates of over 200 ns each for wild type FimH\textsubscript{L}. Due to the tradeoff between the number replicates and simulation length, we performed six replicates of 20 ns of FimH\textsubscript{L} to make comparisons of wild type and mutant FimH, as well as wild type FimH with and without the Cys3-Cys44 disulfide bond.

Since twenty lowest-energy structures are reported for Siglec-8, we performed a single replicate of 50 ns for each structure to compare apo and holo Siglec-8. For the spike RBD-SD1 domains, we performed six replicates of 60 ns.

\subsection*{Backbone and sidechain dihedral angle dynamics}
We used dihedral angles to capture protein dynamics because dihedral angles identify localized regions responsible for the collective displacement of regions distal from the angular rotation, such as in hinged motion \citep{Altis2007DihedralSimulations}. Dihedral angles are also an internal coordinate system that avoids the structure alignment step when using Cartesian coordinates, which can introduce artifacts \citep{Altis2008ConstructionAnalysis}. We use both backbone ($\phi, \psi$) and sidechain ($\chi_1-\chi_5$) dihedral angles. We extracted protein features with MDTraj 1.9.5. 

\subsection*{Inverse of the covariance matrix}
In the literature, the covariance matrix is one approach used to identify protein regions with motions that are related to the motions of many other regions; in particular it is used to identify correlated motions between distant regions in allostery \citep{Melo2020GeneralizedTrajectories, Karami2018InfosteryMutations}. However, constructing networks from the covariance matrix, even with a threshold to remove weak correlations, is susceptible to induced correlations when two nodes (e.g. A and C) are not directly connected but share a connection with a third node (B) \citep{Nguyen2017InverseScience}. Borrowing from the field of network reconstruction, we use the inverse of the covariance matrix to identify the connections and weights, or edges, between nodes \citep{Nguyen2017InverseScience}. This approach is consistent with finding the inverse of a covariance matrix based on C$\alpha$ positions, which fits the Hessian matrix describing an elastic spring network with anisotropy \citep{Atilgan2001AnisotropyModel, Moritsugu2007Coarse-grainedHessian}. We have found that the anisotropic elastic network model (ANM) has large errors when used to describe the motion of FimH, which is consistent with errors for hinge-motion described in literature \citep{Sittel2014PrincipalCoordinates}. As a result, we use dihedral angles. This approach is similar to the torsional network model (TNM) which uses equal spring constants to describe dihedral angles across the protein \citep{Mendez2010TorsionalProteins}. In contrast, the inverse of the covariance matrix uses the variances and the covariances of angles to calculate spring constants for a network of torsional springs. The nodes in our network are dihedral angles, the edges are like linearly coupled torsional springs, and the inverse of the covariance matrix is the Hessian matrix for a TNM. 

To construct our network, we calculate the Moore-Penrose pseudo-inverse of the covariance matrix using both the backbone and sidechain dihedral angles. We use this approach to understand the relative contributions of backbone and sidechain dynamics to collective motion. Since the sign describes whether the angles turn in the same direction, we take the absolute value to get the interaction strength. We do not apply distance filters. While we use the 97\textsuperscript{th} percentile as a value threshold for selecting strong interactions or visualizing the network on the protein, we do not use any thresholds for network comparisons. More generally, we recommend caution for applying thresholds to these networks for analysis. 

\subsection*{Comparing networks of inferred interactions}
To identify interactions that are stronger in one protein state than another, we compare each edge. We select for large differences between groups, relative to the variability within each group. To do this, we filter for differences larger than twice the standard deviation for each group. To compare an edge $e$ between states $a$ and $b$, each with an ensemble of $m$ and $n$ networks, this is 
$|\langle e_a \rangle_m - \langle e_b \rangle_n | > 2 \sigma_{e_a}$ and $ > \sigma_{e_b}$. We apply this rule without determining statistical significance with corrections for multiple comparisons, in order to see the full effects of comparing all interactions on the matrix. In Supplementary Figure \ref{suppfig:WTvsMUT_all}, we also show an example of comparisons without filtering for large differences, in order to illustrate the persistence of the contact-map pattern. We perform the network comparisons in two ways: 1) for every edge on the network (see Supplementary Figure \ref{suppfig:WTvsMUT_all}), 2) accounting for the multi-layer structure of the network by collapsing the backbone-backbone interactions into residue-residue interactions (Figure \ref{fig:fimh}). We analyzed data in python, using SciPy (RRID:SCR\_008058) and custom packages. 

\section*{Results}
\begin{figure}[ht]
    \centering
    \includegraphics[width=\linewidth]{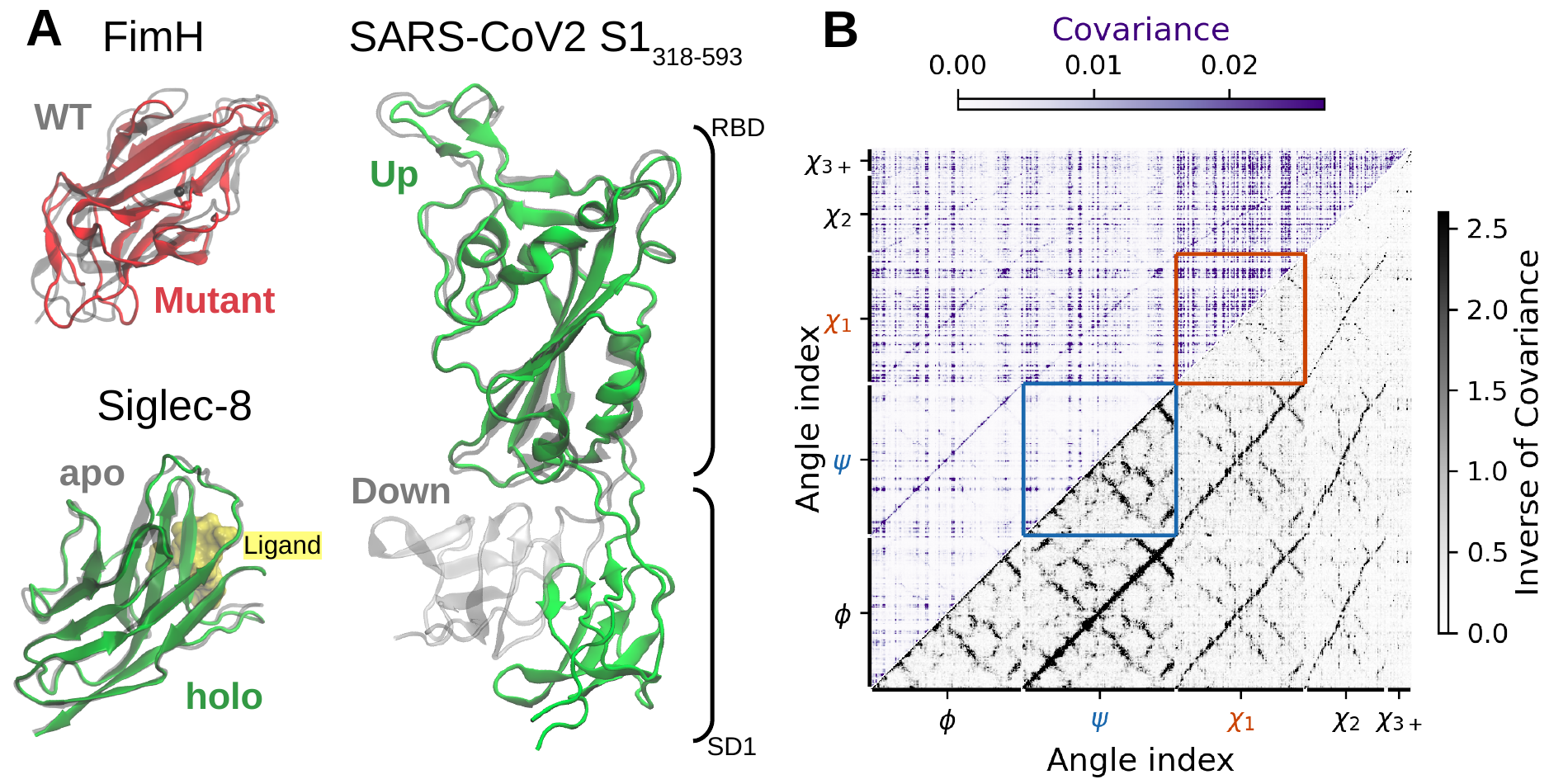}
    \caption{%
    \textbf{Unraveling structural properties from protein conformational dynamics. }
    (\textbf{A}) Cartoon illustrating the three adhesion proteins studied here.
    FimH refers to the lectin domain of a bacterial adhesin found in uropathogenic \textit{E. coli} that binds mannose and undergoes a conformational change under tensile force from urine flow. Siglec-8 refers to the lectin domain of a human immune-inhibitory protein found on eosinophils and mast cells. The SARS-CoV-2 RBD and SD1 domains are thought to undergo a down-to-up transition that makes the RBD available to bind ACE2.
    For each protein, we compare pairs states:
    FimH\textsubscript{L} wild type (PDB 4AUU) and mutant (PDB 5MCA),
    Siglec-8 with ligand 6'S-sLe$^x$ (PDB 2N7B) and without (PDB 2N7A),
    and RBD-SD1 in down and up (PDB 6VSB) states.
    (\textbf{B}) Comparison of covariance analysis of the dynamics (top left) versus the inverse covariance analysis (bottom right) from the dynamics of wild type FimH\textsubscript{L} (see Figure S\ref{suppfig:CvHex_FimH}, S\ref{suppfig:CvHex_Sig8}, and S\ref{suppfig:CvHex_coronavirus} for the other proteins).
    While many studies rely on the analysis of the covariance matrix, our data clearly shows that the structure of the covariance matrix is dominated by artifacts (vertical and horizontal lines) which are stronger for side chain interactions (red square for $\chi_1-\chi_1$). In contrast, the inverse covariance matrix clearly reveals a structure reminiscent of a contact map and is dominated by backbone interactions (blue square for $\psi-\psi$).
    }
    \label{fig:CvH}
\end{figure}

We present results below for these three proteins (Figure \ref{fig:CvH}A). We first focus on the well-characterized allosteric protein FimH\textsubscript{L}. Separation of the FimH\textsubscript{L} domain from its connecting domain (bottom in all figures) is thought to induce an allosteric conformational change on the opposite end of the protein (top in figures) \citep{Trong2010StructuralTwisting, Sauer2016Catch-bondFimH}. This changes the binding pocket from a state with low affinity for the ligand to one with high affinity (Figure \ref{fig:CvH}B) \citep{Trong2010StructuralTwisting}. While wild type FimH\textsubscript{L} is trapped in the high-affinity state, a single-amino acid mutation (Arg60Pro) stabilizes FimH\textsubscript{L} in the low-affinity state \citep{Rodriguez2013AllostericFimH, Rabbani2018ConformationalSystem}. The mutant FimH\textsubscript{L} is of interest because it undergoes an allostery-like conformational change upon binding mannoside ligands and has been proposed as a minimal model of allostery \citep{Rabbani2018ConformationalSystem}.

Like FimH\textsubscript{L}, Siglec-8 binds a carbohydrate ligand and has an immunoglobulin-like fold with two $\beta$-sheets (Figure \ref{fig:CvH}C). Functionally however, Siglec-8 binds to specific sugars found uniquely in human airway tissues to prevent autoimmunity \citep{Propster2016StructuralSiglec-8, Gonzalez-Gil2018SialylatedAirways}. For specific binding, Siglec-8 has a surprisingly rigid binding pocket loop, leading to similar structures for the apo and holo states \citep{Propster2016StructuralSiglec-8}. The SARS-CoV-2 spike protein RBD binds the human ACE2 receptor with a "hook" region\citep{Henderson2020ControllingConformation}. The hook becomes accessible in the up state when the hinge between the RBD and its connector opens (Figure \ref{fig:CvH}A). While the hook and interdomain hinge regions are flexible, the bulk of the RBD is thought to be structurally similar in the up and down states. Both Siglec-8 and the spike RBD-SD1 present challenges for detecting differences in inferred interactions because the protein state changes without major structural changes within domains.

\subsection*{Define and validate network inference from inverse covariance analysis}

Our approach for constructing a network representation of the dynamics of a given protein is comprised of three steps. In the first step, we obtain temporal dynamics for the nodes, which are the backbone ($\phi$, $\psi$) and sidechain ($\chi_{1-5}$) dihedral angles for each residue \citep{DuBay2011Long-rangeAlone, Altis2007DihedralSimulations}. In the second step, we calculate the circular covariance for dihedral angles \citep{Altis2007DihedralSimulations}, which can be thought of as the linearization of the interactions captured by mutual information.
In the third step, we invert the covariance matrix using the Moore-Penrose pseudo-inverse to calculate the best fit for a linear coupling system that can give rise to the observed covariance matrix \citep{Nguyen2017InverseScience}.

Similar to mutual information calculations conducted on other proteins \citep{DuBay2011Long-rangeAlone, Singh2017QuantifyingCARDS}, we find that backbone-backbone interactions computed from the covariance are weak compared to sidechain-sidechain interactions (compare red and blue boxes in Figure \ref{fig:CvH}B and the mutual information in Figure S\ref{suppfig:MI_FimH}). In these networks, some dihedral angles have a banding pattern, suggesting long-range interactions with many other dihedral angles (Figure \ref{fig:CvH}B and Figure S\ref{suppfig:CvHex_FimH}). In contrast, the inverse covariance matrix has localized and specific interactions. In addition, the stronger backbone-backbone interactions have a repeating pattern that resembles the contact map of the protein, and this pattern appears to repeat more weakly in backbone-sidechain interactions.

The banding pattern in the covariance matrix is also widespread in mutual information matrices, where they have been interpreted as long-range interactions important in protein allostery \citep{DuBay2011Long-rangeAlone, Singh2017QuantifyingCARDS}. The large number of long-distance edges produce "hairball" networks, which led to the use of pruning algorithms \citep{Cruz2020DiscoveryExperiments, Knoverek2021OpeningActivity} or distance filters \citep{Melo2020GeneralizedTrajectories, Karami2018InfosteryMutations, Grant2021TheBioinformatics} in prior studies, in order to make network analysis tractable. Thus, we wondered whether the long-range interactions are capturing a physical feature of the dynamics. To answer this question, we investigate the reproducibility of the covariance, correlation, and inverse covariance matrices extracted from different replicates of MD simulations.

\begin{figure}[h]
    \centering
    \includegraphics[width=\linewidth]{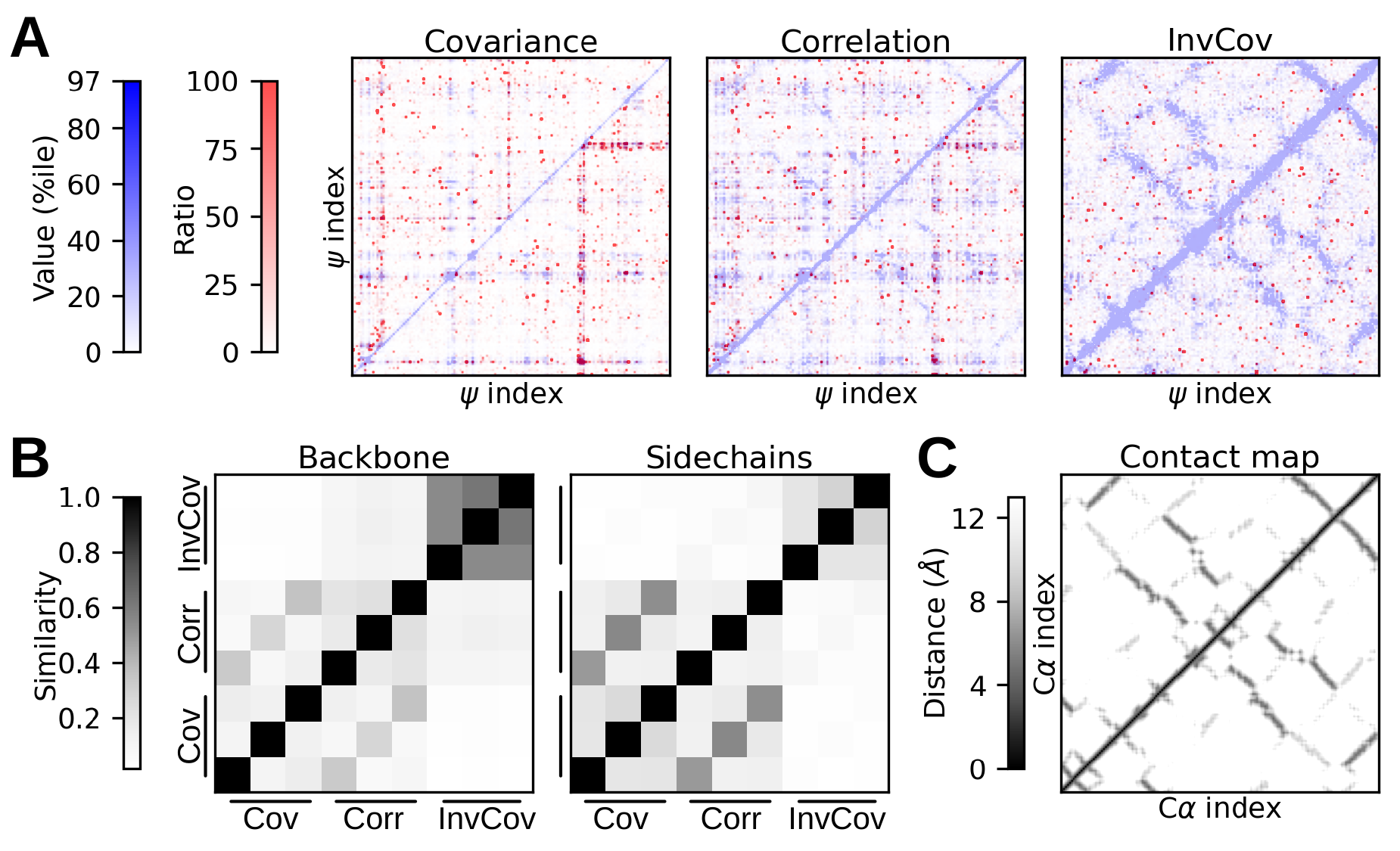}
    \caption{%
    \textbf{ Inverse covariance matrix is robust across replicates whereas covariance and correlation matrices are not.}
    (\textbf{A}) Triangular regions above and below the matrix diagonal show results from two replicates of wild type FimH\textsubscript{L} starting from the same protein structure. We show $\psi-\psi$ interactions. We show interaction strength in blue with a normalization for each triangular region made based on the 97\textsuperscript{th} percentile of observed strengths. In red, we show the ratio for interaction strength between the two replicates. Purple indicates strong interactions that are not reproduced in the other replicates. (See Figure S\ref{suppfig:similar_95thpile} for weaker interactions visible when normalized to the 95\textsuperscript{th} percentile for $\psi-\psi$ and $\chi_1-\chi_1$.)
    For covariance and correlation matrices, we find that backbone-backbone interactions are mostly quite weak, but the strong interactions (darker blue) vary drastically between replicates.  In contrast, for the inverse covariance matrix, the strongest backbone-backbone are symmetric across the diagonal.
    (\textbf{B}) To evaluate the robustness of network inference, we calculate the Jaccard similarity coefficient for the covariance, correlation, and inverse covariance analyses methods across three simulation replicates. We define edges above the threshold of $\geq$97\textsuperscript{th} percentile. (See Figure S\ref{suppfig:similar_jaccard} for other thresholds.) In grey scale, we show similarity separately for $\psi-\psi$ and $\chi_1-\chi_1$ interactions. Darker grey indicates results are similar across replicates for the inverse covariance approach and much less similar for the other two methods.
    (\textbf{C}) The inverse covariance matrix resembles the contact map. We show the C$\alpha$ inter-residue distance from the crystal structure. Darker grey indicates shorter distance.
    }
    \label{fig:similar}
\end{figure}

In Figure \ref{fig:similar}A, we contrast the matrix for one replicate in the upper-diagonal with the second replicate in the lower-diagonal and quantify the similarity in Figure \ref{fig:similar}B. For replicate MD simulations, we used the same initial protein structure with randomized solvation and initial velocities. Both covariance matrices have banding patterns suggesting hotspots that interact with many residues across the protein. However, each replicate has its own banding pattern, with interaction strengths that are over ten times greater than those found in the other replicate, indicating high variability in the networks that one would construct from replicate simulations.

\FloatBarrier
Since the banding pattern is associated with dihedral angles with high variance, we also consider the correlation matrix, which normalizes the covariance matrix by the variance of each dihedral angle. It is visually apparent that the banding in the covariance matrix is not simply due to high variance because there are still bands in the correlation matrix.  While normalizing to the correlation matrix uncovers some interactions in a contact map pattern, they are weak compared to the banding pattern (see Figure S\ref{suppfig:similar_95thpile} for a lower maximum value on the color map scale).

In contrast to the irreproducible results obtained with the covariance matrix, for the inverse of the covariance matrix, we find a pattern that is visually similar to the 12\AA\ contact map (Figure \ref{fig:similar}C). The diagonally symmetric contact map pattern in blue indicates similarly strong interactions for two replicates (Figure \ref{fig:similar}A). After quantifying the robustness across three replicates using the Jaccard similarity index, we find that the inverse covariance has higher similarity (59-72\% shared edges) than the covariance (8-10\%) or the correlation (13-16\%) for $\psi-\psi$ backbone interactions (Figure \ref{fig:similar}B). The $\chi_1-\chi_1$ similarity values for inverse covariance are lower, but still higher than for the other two methods.
These data clearly demonstrate that networks inferred from inverse covariance analysis are more robust than networks constructed from correlational measures.

\FloatBarrier
\subsection*{Inverse covariance analysis yields structural networks}

\begin{figure}[]
    \centering
    \includegraphics[width=\linewidth]{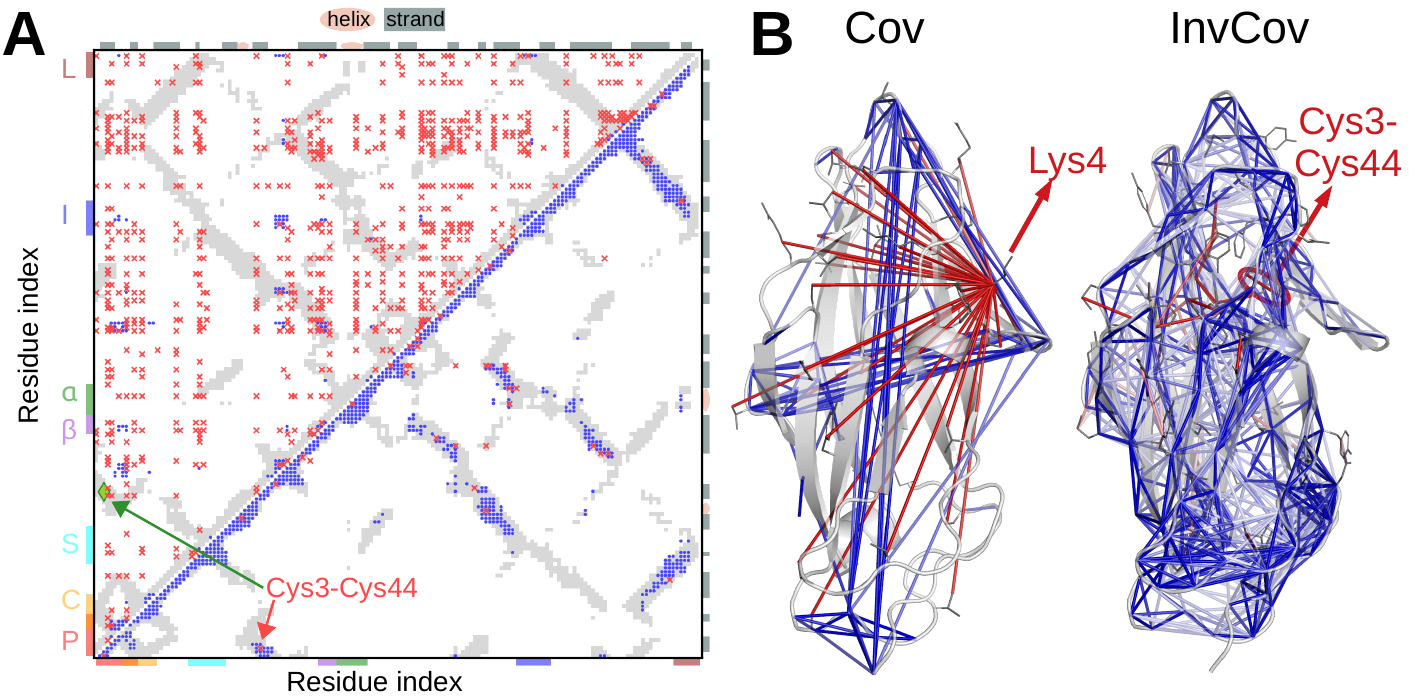}
    \caption{\textbf{The inverse covariance matrix enables us to extract a ``contact map''-like network from the protein dynamics.}
    (\textbf{A}) Comparison of strong interactions identified for the covariance matrix (top left) and for the inverse covariance matrix (bottom right) of wild type FimH.
    To provide context for our data, we plot the 12 \AA\ contact map in grey within the matrix. On the top and right axes we show helix (pink) and strand (teal) secondary structures assigned by the Dictionary of Secondary Structure of Proteins algorithm (DSSP) . On the left and bottom axes we show putative allosteric pathway landmarks \citep{Rodriguez2013AllostericFimH}: pocket zipper (red), clamp segment (yellow), swing loop (cyan), $\beta$-bulge (purple), $\alpha$-switch (green), insertion loop (blue), and linker loop (dark red).
    We only show strong edges, with the threshold set at the 97\textsuperscript{th} percentile of all dihedral interactions. See Figure S\ref{suppfig:net_thresh} for other cutoffs. We averaged edge weight across three replicates. The blue dots represent the average of backbone-backbone interactions by residue. The red crosses represent sidechain-sidechain interactions ($\chi_1-\chi_1$).
    The inverse covariance network are predominantly backbone interactions that fall within the 12 \AA\ contact map. There is a $\geq$ 99\textsuperscript{th} percentile $\chi_1-\chi_1$ interaction the Cys3-Cys44 disulfide bond (red arrow). However, this interaction is only 80\textsuperscript{th} percentile in strength (green diamond and arrow) for the covariance matrix, which is dominated by other $\chi_1-\chi_1$ interactions.
    (\textbf{B}) Since the covariance matrix has many long-range interactions, we only show the backbone interactions and the sidechain interactions for Lys4. 
    In contrast, the inverse covariance network has mostly short-range interactions, including the disulfide bond. We show backbone interactions on the C$\alpha$ atoms and sidechain interactions on the the 4\textsuperscript{th} $\chi_1$ atoms. 
    We show backbone interactions in blue and sidechain interactions in red. Darker colors indicate stronger interactions.
    }
    \label{fig:net}
\end{figure}

Prompted by the strong visual resemblance between the inverse covariance network and the 12\AA\ contact map and the complete absence of this pattern in the covariance network, we wondered if the inverse covariance matrix could be used to identify specific physical interactions. To answer this question, we overlaid the strongest edges ($\geq$ 97\textsuperscript{th} percentile) on the 12\AA\ contact map, highlighting the backbone-backbone edges as blue dots and the sidechain-sidechain edges as red crosses (Figure \ref{fig:net}A and Figure S\ref{suppfig:multi_pattern}).

To correct for the high variability across replicates we previously found for the covariance networks, we averaged networks across the three replicates. Despite the averaging, the covariance network shows strong interactions across distant protein regions and are dominated by $\chi_1-\chi_1$ interactions.

\FloatBarrier
To understand how these contrasting patterns affect interpretation, we next visualize strong interactions as edges drawn on the protein structure (Figure \ref{fig:net}B). Since drawing all edges would make the covariance network indecipherable, we only show the edges originating from Lys4. 
In contrast, the inverse covariance network uncovers edges that mostly connect physically close residues. Specifically, we find a $\chi_1-\chi_1$ edge between Cys3-Cys44 for the sole disulfide bond in FimH\textsubscript{L}, and that this edge is missing from the covariance network (red vs green arrow in Figure \ref{fig:net}A).

Examining the contact map pattern of the inverse covariance network in more detail, we compare edge weight with the distance between C$\alpha$ atoms (Figure S\ref{suppfig:multi_edge}). We find that for backbone-backbone interactions, the strongest interactions are between residues connected by a peptide bond, followed by hydrogen bonds within $\beta$ sheets, and then non-bonding interactions.

After examining backbone-backbone interactions, we next looked at the progressively weaker interactions involving sidechains distal from the backbone (Figure \ref{fig:CvH}B and Figure S\ref{suppfig:multi_pattern}).
We find the contact map pattern is still apparent for $\phi-\chi_1$ or $\psi-\chi_1$ interactions, but becomes very weak for  $\phi-\chi_2$ or $\psi-\chi_2$ interactions, and becomes indistinguishable from noise for interactions between proximal and distal sidechain dihedrals. The inverse covariance analysis thus suggests that backbone dihedral motion is most strongly coupled to nearby backbone dihedrals and has more dissipated effects on sidechain dihedrals. This relationship is consistent with how backbone motions can sterically trap or free sidechains, whereas sidechain motions are more limited in their impact on backbone motion \citep{Keedy2012TheMutations}.

The different strengths of interactions for backbone-backbone and backbone-sidechain edges suggests that qualitatively different types of interactions have different properties. For two residues \textit{i} and \textit{j}, the backbone-backbone edge $\phi-\psi[i, j]$ are larger than the backbone-sidechain edge $\chi_1-\chi_1[i,j]$, which is consistent with the physical differences between these two edge types. Moreover, most backbone-backbone edges within the same residue, $\phi-\psi[i, i]$, are stronger than backbone edges connecting to other residues, $\phi-\phi$[i, j] and $\psi-\psi$[i, j] (Figure S\ref{suppfig:multi_edge}).

\FloatBarrier
\subsection*{Detecting both large and small structural changes in FimH\textsubscript{L}}
\subsubsection*{Conformational differences between wild type and mutant}
As a way to further validate our approach, we next test if we are able to identify the well-characterized differences between wild type FimH\textsubscript{L} and the Arg60Pro mutant. To compare inferred networks for the wild type and mutant proteins, we identified edges where the average difference was larger than two times the standard deviation across each group of replicates. We performed this analysis once with the entire set of edges (Figure S\ref{suppfig:WTvsMUT_all}), and again with only the backbone interactions collapsed into a residue interaction network. For our comparisons and the matrix visualization of the differences, we do not apply a distance filter or a threshold for the edge-strength. However, for the visualization on the protein, we only show differences with magnitude larger than the 97\textsuperscript{th} percentile.

\begin{figure}[]
    \centering
    \includegraphics[width=1\linewidth]{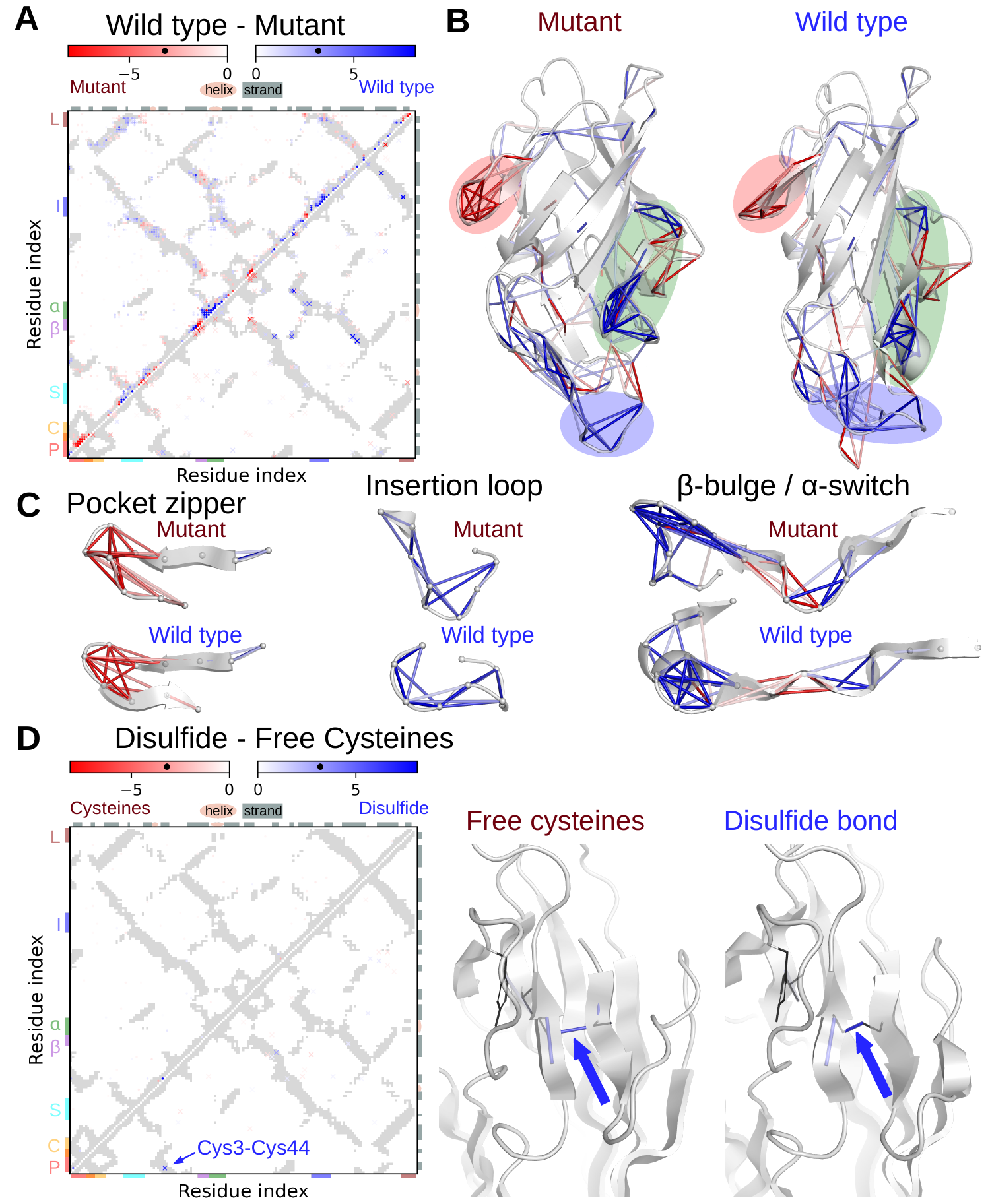}
    \caption{\textbf{Inverse covariance analysis can detect both large and small structural changes in FimH\textsubscript{L}.
    }
    (\textbf{A}) Comparing inferred networks for wild type and mutant proteins, we show differences in the backbone (top left in dots) and $\chi_1-\chi_1$ (bottom right in crosses). 
    As in Figure \ref{fig:net}A, we annotate the 12 \AA\ distance cutoff, secondary structures, and landmarks. The black dots on the colorbar mark the 97\textsuperscript{th} percentile in magnitude. On the adjacency matrix, we show all differences greater than 2$\sigma$. 
    (\textbf{B}) On the protein, we show large differences ($\geq$ 97\textsuperscript{th} \%ile). Red shows interactions stronger for mutant fimH\textsubscript{L}; blue for WT fimH\textsubscript{L}. We highlight the pocket zipper (red), insertion loop (blue), and $\beta$-bulge/$\alpha$-switch (green) and 
    (\textbf{C}) show isolated parts of the network. 
    (\textbf{D}) Comparing wild type FimH with the Cys3-Cys44 disulfide bond intact or reduced \textit{in silico}.
    (\textbf{E}) The blue lines shows that the Cys3-Cys44 $\chi_1-\chi_1$ (blue arrow) and the Phe43-Cys44 backbone-backbone interactions are stronger when the disulfide bond is intact.
    }
    \label{fig:fimh}
\end{figure}

The visualization of the differences enables us to identify several interactions stronger in either the mutant or the wild type proteins (red or blue patches, respectively, in Figure \ref{fig:fimh}A). This is consistent with the difference in initial structure (RMSD=3.15\AA), dihedral dynamics \citep{Liu2021ConformationalMutation}, and the residues that are rearranged in the allosteric conformational change \citep{Rodriguez2013AllostericFimH, Kisiela2021ToggleCore}.

\FloatBarrier
For concreteness, we focus on two regions at the edge of the protein structure that are easier to visualize: the binding pocket zipper at the top of FimH\textsubscript{L}, and the insertion loop at the bottom.
In the pocket zipper, we found much stronger interactions for the mutant protein (median:2.8-fold, IQR:2.0--5.2-fold), which correspond to smaller dihedral fluctuations \citep{Liu2021ConformationalMutation}.
On the other hand, in the insertion loop, we identified changes in interaction that were stronger in the wild type than the mutant protein (3.2-fold, 2.2--3.7-fold). Structurally, this is consistent with how the insertion loop is stabilized in the wild type structure and exposed to solvent in the mutant protein \citep{Interlandi2016MechanismSimulations, Rabbani2018ConformationalSystem}. Dynamically, stronger interactions within the insertion loop is consistent with smaller dihedral fluctuations in the wild type protein \citep{Liu2021ConformationalMutation}.

We further identify differences at the $\beta$-bulge, $\alpha$-switch, and swing loop regions of the allosteric pathway, consistent with structural differences between wild type and mutant proteins. The mutant protein has stronger interactions within the loop formed by the $\beta$-bulge (2.0-fold, 1.2--2.3-fold) and also with a nearby loop. In the wild type protein, the loop is smoothed out into a $\beta$-strand. The wild type protein has stronger interactions (1.8-fold, 1.5--2.6-fold) in the $\alpha$-helix, compared to the $3_{10}$-helix in the mutant protein, which is probably due to different hydrogen bonding patterns. In the swing loop, we again find stronger interactions in the wild type protein (2.0-fold, 1.6--2.7-fold). 

For these allosteric pathway landmarks, it is visually apparent that we detect large differences in inferred interactions when structures are closer in one state and stretched apart in the other state. Beyond these regions, there are several other regions with similarly large changes in interaction between the wild type and mutant proteins, shown in blue and red patches (Figure \ref{fig:fimh}A).

\subsubsection*{Quantifying the impact of disulfide bond reduction}
We next used wild type FimH\textsubscript{L} to explore the impact of reducing the single disulfide bond between Cys3-Cys44 \textit{in silico} on fast, nanosecond-timescale dynamics. Using the inverse covariance analysis, we correctly identified the 6-fold stronger $\chi_1$-$\chi_1$ interactions in the presence of the disulfide bond, which was the largest difference detected (Figure \ref{fig:fimh}D). This matches our expectations because the covalent bond between the most distal atoms forming the $\chi_1$ rotamer angle directly couples $\chi_1$ dynamics. Together, these analyses show that the inverse covariance analysis method is sensitive to both local differences and conformational differences.

\FloatBarrier
\subsection*{Network inference detects key mechanisms involved in Siglec-8 binding}
Like FimH\textsubscript{L}, the human immune cell adhesion protein, Siglec-8, is also a lectin with an immunoglobulin-like fold with a single disulfide bond. For Siglec-8, we compare the apo (no ligand) and holo (bound to the native 6'S-sLe\textsubscript{x} ligand) states. Due to the rigid binding pocket loop that only differs by a few sidechain rearrangements, apo and holo Siglec-8 have extremely similar structures\citep{Propster2016StructuralSiglec-8}. Intriguingly, the rigidity of the CC' binding pocket loop in apo Siglec-8 occurs in the absence of stabilizing secondary structure motifs \citep{Propster2016StructuralSiglec-8} and plays a major role in recognizing specific ligands in the airway to avoid autoimmunity \citep{Gonzalez-Gil2018SialylatedAirways}. 

\begin{figure}[]
    \centering
    \includegraphics[width=1\linewidth]{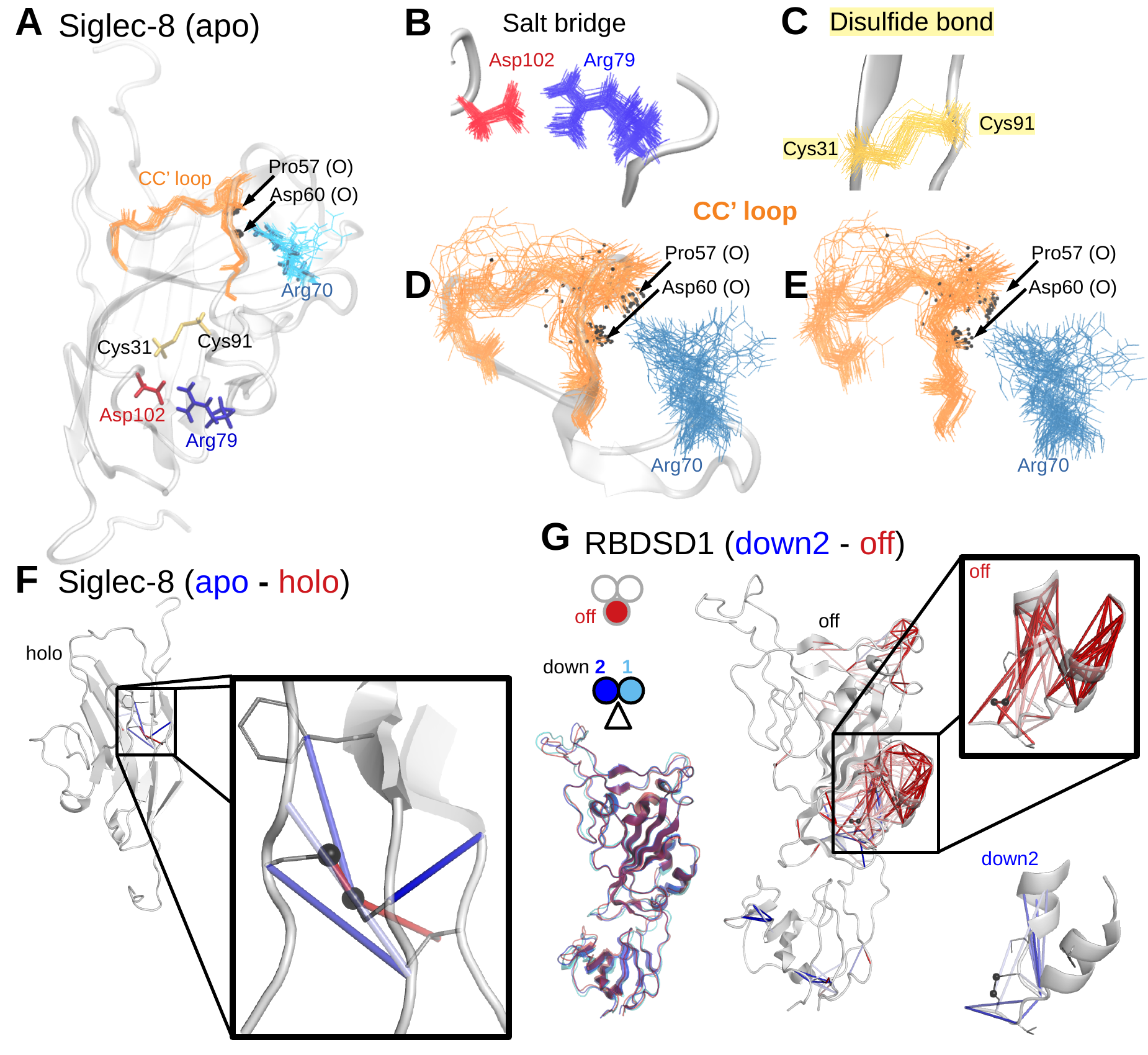}
    \caption{\textbf{Inferred networks identify strong interactions and changes in interaction strength.}
    (\textbf{A}) Illustration of Siglec-8 highlighting the CC' loop of the binding pocket in orange, Arg70 in cyan, and hypothesized hydrogen bond acceptor carbonyl oxygens (black spheres) for ensemble of 20 NMR structures. 
    From MD simulations of all structures, we show snapshots every 10ns for the 
    (\textbf{B}) Cys31-Cs91 disulfide bond, 
    (\textbf{C}) Arg79-Asp102 salt bridge, and 
    (\textbf{D}) CC' loop.
    Structures were aligned to the backbone atoms for residues 13-135 to exclude the N- and C-terminus tails. These snapshots show the stability of the salt bridge and disulfide bond, and the surprising stability of the unstructured CC' loop. 
    However, the hypothesized interaction between Arg70 and the CC' loop is much less stable, even with 
    (\textbf{E}) structural alignment using the backbone of the CC' loop. 
    (\textbf{F}) Beyond identifying strong interactions, our approach also identifies rearrangements of strong interactions without large structural changes in Siglec-8 and the SARS-CoV-2 RBD-SD1.
    Siglec-8 with (holo) and without (apo) ligand have similar structures. However, inverse covariance analysis reveals rearrangement of strong interactions in a region opposite the binding pocket, including the Cys31-Cys91 disulfide bond (black spheres). We show interactions stronger in holo (red) and apo (blue) on the holo structure. 
    (\textbf{G}) For the SARS-CoV-2 spike protein, rotation around the RBD-SD1 hinge exposes the RBD in the up conformation (Figure \ref{fig:CvH}A), 
    while the structures with the RBD hidden are extremely similar (down1, down2, and off conformation). 
    Comparing inferred network interactions, we detect differences near the hinge, including the Cys336-Cys361 disulfide bond (black spheres) and nearby $\alpha$-helices. For these regions, we show the interactions that are stronger in the down2 (blue) and off (red) conformations on the corresponding structure. 
    See Figure S\ref{suppfig:RBDSD1all} for other comparisons.
    }
    \label{fig:combo}
\end{figure}

One hypothesized mechanism for stabilizing the CC' loop in apo Siglec-8 is that the Arg70 sidechain forms hydrogen bonds with the loop backbone at Pro57 and Asp60 (Figure \ref{fig:combo}A)\citep{Propster2016StructuralSiglec-8}. 
We identified Arg70-Pro57 or Arg70-Asp60 hydrogen bonds with occupancy between 10-33\% in seven out of 20 simulations from the ensemble of NMR structures, and higher occupancy (80 and 83\%) in only two simulations. This suggests the hydrogen bond is rarely present. Nonetheless, the unstructured CC' loop has surprisingly small fluctuations across the ensemble of MD simulations (Figure \ref{fig:combo}B-E). 

\FloatBarrier
Using our inferred interaction networks, we identified strong interactions within the CC' loop, as well as interactions from outside the loop to its hinges at Ala53 and Pro62 (Figure S\ref{suppfig:sig8CCloophbond}). Consistent with the hydrogen bonding analysis, we only identified strong interactions with Arg70 in a few MD simulations, which disappeared after averaging the networks across the ensemble. Counter-intuitively, in the two structures where Arg70 does form stable hydrogen bonds with Pro57 and Asp60, the CC' loop has paradoxically larger fluctuations, especially at Tyr58 and Gln59 at the loop tip (Figure S\ref{suppfig:sig8CCloophbond}). Taken together, we suspect that steric hindrance plays a role in stabilizing the CC' loop internally and externally at the hinge edges. 

Of note, this example demonstrates an instance where the inferred network approach provides more clarity than contact or interaction energy networks, because the inverse of the covariance matrix identifies the degrees of freedom that most strongly affect the fluctuations at the CC' loop. 

After examining why the CC' loop is similar in apo and holo Siglec-8, we next compared inferred networks for the apo and holo states. The holo state has stronger $\chi_1$-$\chi_1$ interactions corresponding to the Cys31-Cys91 disulfide bond (Figure \ref{fig:combo}C and Figure S\ref{suppfig:siglecall}). The disulfide bond is conserved in the siglec family and is located on the sheet of the $\beta$-sandwich opposite the binding pocket \citep{Propster2016StructuralSiglec-8}. Nearby, we also observe other changes in interaction strength involving Asp90. Although distant from the binding site, the Asp90-Cys91-Ser92 motif in Siglec-8 is a variant of the Asn-Cys-Ser or -Thr motif that is conserved in the rest of the siglec family \citep{Freeman2001ARecognition}. Differences in interaction strength between apo and holo states identify changes in the dynamics of this evolutionarily conserved region during ligand-binding.

In contrast, reducing the Cys31-Cys91 disulfide bond in the holo state \textit{in silico} has a different pattern (Figure S\ref{suppfig:siglecall}). Both ligand-binding and the presence of the disulfide bond correspond to stronger a $\chi_1$-$\chi_1$ interaction at Cys31-Cys91. Both conditions also stabilize Cys31, indicated by decreased backbone dihedral fluctuations and increased duration within an extended secondary structure as assigned by the DSSP algorithm, and shorter Cys31-Cys91 C$\alpha$ distance. Taken together, we find that the network rearrangement that occurs with ligand-binding increases Siglec-8 conformational stability near an evolutionarily conserved disulfide bond, even though the region is not near the binding site.

\subsection*{Comparison of the spike protein RBD-SD1 fragments in the up, down, and off states shows network rearrangements without large structural differences}
Next, we investigated whether there are differences in the networks inferred from the `up', `down', and `off' states of the RBD-SD1 domains of the SARS-CoV-2 spike protein (Figure \ref{fig:CvH}A). The RBD connects to SD1 (Fig \ref{fig:combo}E) via two hinge-like loops that are more flexible in the up state than in the down or off states \citep{Ke2020StructuresVirions}. While the up state has a different orientation around the hinge than the down and off states, they all have similar SD1 structures (C$\alpha$-RMSD $\leq$ 0.64\AA) and somewhat similar RBD structures (C$\alpha$-RMSD $\leq$ 1.55\AA). 

Opening the hinge angle in the down-to-up transition is thought to make the binding site on the RBD available to attach to human ACE2 \citep{Wrapp2020Cryo-EMConformation, Henderson2020ControllingConformation}. To focus on the RBD-SD1 hinge, we isolated these domains from the rest of the spike protein and ignored glycosylated sugars. While these simplifications limit the strength of our conclusions into the function of the spike protein, the RBD-SD1 structures nonetheless provide a useful system for comparing dynamics in a system that initially resembles rigid-body motion around a hinge.

Since the trimeric spike protein with one exposed RBD has two hidden RBDs, we isolated one RBD-SD1 fragment in the `up' state, and two fragments in the `down' state. We obtained another RBD-SD1 domain in the `off' state from a structure with all RBDs hidden and 3-fold rotational symmetry. We first compared the RBD-SD1 fragments in the down and off conformations. Despite the structural similarity, we nevertheless detect differences in the inferred networks (Fig \ref{fig:combo}E) based on dynamics. We find the networks for the two down conformations are more similar to each other than to the networks for the off conformation (Figure S\ref{suppfig:RBDSD1all}).  

We next compared RBD-SD1 fragments in the down and off conformations to the fragment in the up conformation (Figure S\ref{suppfig:RBDSD1all}). Surprisingly, we find that the differences in inferred networks among the down and up conformations are more localized, while the networks for the off conformation are distinct from the others. The localized differences are in the hook of the RBD that is exposed in the up conformation, and at an $\alpha$-helix near the RBD-SD1 hinge. Intriguingly, we also identified differences in the structural orientation and the inferred network interaction at the Cys336-Cys361 disulfide bond. In this region, one down conformation is more similar to the up conformation, while the other down conformation resembles the off conformation (Fig \ref{fig:combo}E). 

As a proof-of-concept, we investigated whether it is possible to infer an interaction network for the trimeric SARS-CoV-2 spike protein, even though it is much larger than the RBD-SD1 domains. However, the amount of data required for network inference using the inverse covariance method scales faster than the number of residues. As a result, our approach required more data than was available from the two sets of publicly-accessible simulations for which the up and down states are labeled \citep{Casalino2020BeyondProtein, ResearchMolecularSARS-CoV-2}. Instead, we chose the longer simulations that explore the transition among the down, up, and open states published by the Bowman lab \citep{Zimmerman2021SARS-CoV-2Proteome}. Using 100,000 snapshots, we were able to infer an interaction network for the 3,363 residues of the spike protein (Figure S\ref{suppfig:spike_H}). The number of snapshots is an order of magnitude larger than those currently available for labeled states. Thus, it may be feasible to compare states for the entire S1/S2 complex of the spike protein if there is sufficient data, or by choosing a less data-intensive method for solving the inverse problem.

\section*{Discussion}
We identify some of the shortcomings of correlation-based approaches for network inference from protein dynamics using the covariance, correlation, and mutual information matrices. These networks have low reproducibility among replicates and exhibit long-range connections that are difficult to tie to physical explanations. To address these shortcomings, we use the inverse of the covariance matrix. This is a well-established technique from network inference\citep{Nguyen2017InverseScience} for solving the inverse problem for a system that can produce the observed correlated dynamics. Our approach builds networks where each node is a dihedral angle, including both the backbone ($\phi$, $\psi$) and the sidechains ($\chi_{1-5}$), and edges are inferred from the coupling interactions between angles. We chose the internal coordinate system of dihedral angles\citep{Altis2007DihedralSimulations} to easily include sidechain dynamics, localize hinges that drive distal dynamics\citep{Liu2021ConformationalMutation}, and to avoid the alignment step in Cartesian coordinates that introduces `artifacts' in hinged motion \citep{Sittel2014PrincipalCoordinates}. 

Using the inverse covariance approach, we detected differences in conformation, subtle differences between protein states without large conformational changes, and localized perturbations in the structure of biomedically important proteins. The inverse covariance networks capture a hierarchy of interactions that resemble that qualitatively different types of interactions, suggesting a multi-layer network structure. The strongest edges connect dihedral angles with covalently-bonded atoms, with weaker interactions for greater distances. Moreover, the contact map-like pattern found in backbone-backbone interactions are repeated more weakly in backbone-sidechain and sidechain-sidechain interactions. This hierarchy of inferred interactions is consistent with the view that smaller backbone rearrangements related to larger sidechain motions \citep{Keedy2012TheMutations}. 

Our results suggest that solving the inverse problem uncovers the underlying interactions that ultimately drive protein dynamics, but are not well-captured by cataloguing the observed correlated motions or comparing static structures. Notably, inverting the covariance matrix is the simplest of a variety of tools available for network inference from dynamics used in network science \citep{Nguyen2017InverseScience, Peixoto2019BayesianBlockmodeling}. While the simplicity of inverting the covariance matrix increases accessibility, there are some obvious limitations \citep{Nguyen2017InverseScience}. We calculate the circular covariance matrix on dihedral angle distributions that are multi-modal. The inverse of the covariance matrix is analogous to linearly coupled torsional springs, which do not represent the complexity of atomic interactions within a protein. Moreover, network inference by inverting the covariance matrix requires a large amount of data \citep{Nguyen2017InverseScience}. Our work establishes a baseline approach, which can be easily built upon by incorporating more sophisticated \citep{Nguyen2017InverseScience, Wang2018InferringNetworks} -- and yet more involved -- approaches that better describe dihedral distributions \citep{Marks2011ProteinVariation} or account for nonlinear interactions \citep{Banerjee2019UsingLinks}. 

Despite these limitations, our approach yielded significant insights for three adhesion proteins. Comparing the networks inferred for two protein states at a time, we were able to tie differences in inferred network structure to structural and dynamical differences. For FimH\textsubscript{L}, a comparison of inferred networks for the wild type and mutant proteins identifies protein regions with conformational changes consistent with the allosteric pathway sites \citep{Rodriguez2013AllostericFimH}.
For Siglec-8 and the SARS-CoV-2 RBD-SD1 construct, we were able to detect network rearrangements despite the similar structures of Siglec-8 in the apo and holo states, and of the individual RBD and SD1 domains in the up and down states. In Siglec-8, we were also able to use strong interactions identified by the network to propose a new mechanism for stabilizing an unstructured loop in the binding pocket. This serves as an example where the network inference approach has an advantage over contact and interaction energy networks. Taken together, our results show that the network inference approach can identify protein regions of interest based on dynamical differences that are rooted in physically interpretable interactions. 

\FloatBarrier
\bibliography{references}

\section*{Supporting Information}
\subsection*{Network inference from the inverse covariance matrix of dihedral angles}
Using dihedral angles enables us to easily account for sidechain dynamics avoid the artifacts introduced by structural alignment \cite{Altis2007DihedralSimulations}. While protein dihedral angles have steep energy wells, we make the simplifying assumption that each dihedral angle has a harmonic potential energy function centered around one preferred orientation. This is only reasonable for short, picosecond-timescales, which is usually faster than the timescale for backbone dihedral flipping \cite{Zwier2010ReachingSimulations}. Moreover, let's assume that dihedral angle pairs only have linear coupling, which neglects nonlinear interactions and many-body interactions\cite{Altis2008ConstructionAnalysis, Sittel2014PrincipalCoordinates,Mendez2010TorsionalProteins}. 

We can describe these linearly coupled dihedral angles as the Hessian matrix, $\textbf{H}$, where $H_{ij}$ is the coupling constant between a pair of dihedrals, $i, j$. A positive coupling indicates rotation in the same direction. A negative coupling indicates rotation in opposite directions.

The harmonic coupling assumption means that $H_{ij}$  goes as $\frac{energy}{radians^2}$. Radians are a unitless measure, so $H_{ij}$ has units $J=Nm$, but we find it easier to use $J/rad^2$. Some combination of angular displacements from equilibrium, $\theta$, then produces the torque vector $f=-\textbf{H} \theta$. The potential energy difference from that at equilibrium is $\Delta U = \int - f d\theta = \frac{1}{2} \theta^T \textbf{H} \theta$.

We can also the use the Boltzmann relationship to describe the probability of a particular configuration in terms of angular displacements, $\theta$, as $p[\theta] \sim exp[- \Delta U / k_b T]$. The exponent is $-\theta^T \textbf{H} \theta / (2 k_b T)$ and can be rearranged to $-\frac{1}{2} \theta^T (\frac{1}{k_b T} \textbf{H}) \theta$. This yields the probability distribution function in terms of $\textbf{H}$ (equation \ref{eq:boltzmann}).

\begin{equation}
\label{eq:boltzmann}
p[\theta] = exp[-\frac{1}{2} \theta^T (\frac{1}{k_b T} \textbf{H}) \theta]
\end{equation}

By recognizing the form of a multivariate Gaussian distribution with covariance matrix $\textbf{C}$ (equation \ref{eq:Gaussian}), we can see the inverse relationship between the Hessian and covariance matrices (\ref{eq:invcov}). This derivation mirrors the one for anisotropic elastic network models \cite{Atilgan2001AnisotropyModel, Moritsugu2007Coarse-grainedHessian}, which use displacements in Cartesian space, instead of an internal coordinate system of dihedral angles. 

\begin{equation}
\label{eq:Gaussian}
p[\theta] = exp[-\frac{1}{2} \theta^T \textbf{C}^{-1} \theta]
\end{equation}

\begin{equation}
\label{eq:invcov}
\frac{1}{k_b T} \textbf{H} = \textbf{C}^{-1}
\end{equation}

\subsection*{Selected examples of the covariance matrix and its inverse}
To demonstrate that the different patterns found between the two matrices apply to multiple MD trajectories, we show the covariance matrix diagonally across from its inverse for FimH (Supplementary Figure \ref{suppfig:CvHex_FimH}), Siglec-8 (Supplementary Figure \ref{suppfig:CvHex_Sig8}), and the SARS-CoV-2 spike protein RBD-SD1 domains (Supplementary Figure \ref{suppfig:CvHex_coronavirus}). For each, we set the color scale maximum to the 97\textsuperscript{th} percentile. We show these in the same format as in Figure \ref{fig:CvH}. 

We highlight the backbone-backbone $\psi$-$\psi$ interaction in blue, and the sidechain-sidechain $\chi_1$-$\chi_1$ interaction in red. The covariance and mutual information (Supplementary Figure \ref{suppfig:MI_FimH}) matrices have stronger $\chi_1$-$\chi_1$ interactions than $\psi$-$\psi$. This is reversed for inverse covariance matrices. 


\begin{suppfigure}
    \centering
    \includegraphics[width=\linewidth]{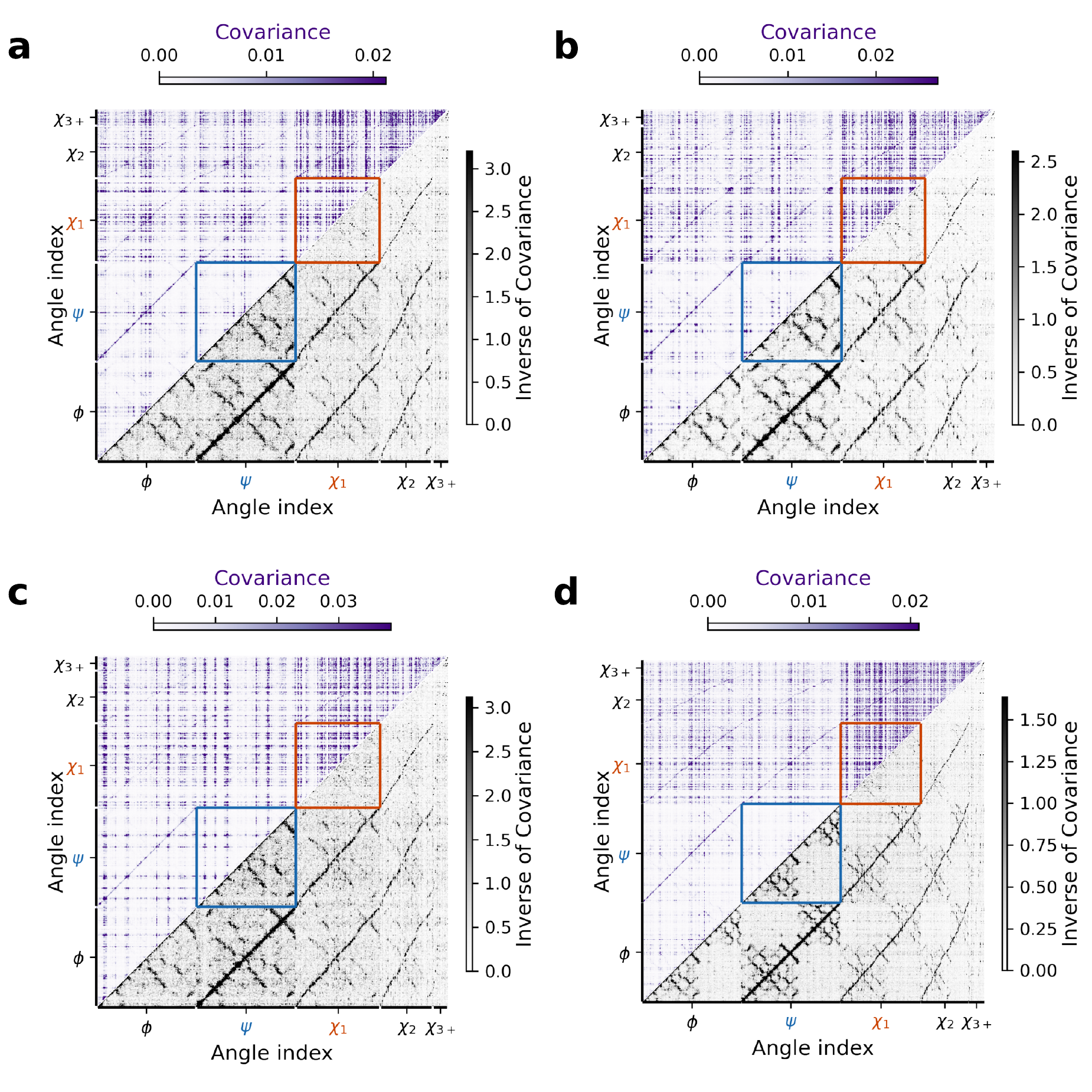}
    \caption{%
    \textbf{The covariance matrix and its inverse show different patterns for FimH.} We show that the inverse covariance matrix resembles the contact map for four representative examples. 
    \textbf{a}, Wild-type FimH\textsubscript{L} (PDBID 4AUU) in 20ns and
    \textbf{b}, 200ns simulations for a different replicate than the one in Figure \ref{fig:CvH}B.
    \textbf{c}, Arg60Pro mutant FimH\textsubscript{L} (PDBID 5MCA) in a 20ns simulation. 
    \textbf{d}, FimH\textsubscript{2} (PDBID 4XOD), which has both the lectin and pilin domains in a 200ns simulation. The two-domain structure is visually apparent in the inverse covariance matrix.
    As in Figure \ref{fig:CvH}B, we show the covariance matrix in the top left triangle and the inverse of the covariance matrix in the bottom right triangle. For each dataset, we set the colorscale maximum to the 97\textsuperscript{th} percentile. 
    }
    \label{suppfig:CvHex_FimH}
\end{suppfigure}

\begin{suppfigure}
    \centering
    \includegraphics[width=\linewidth]{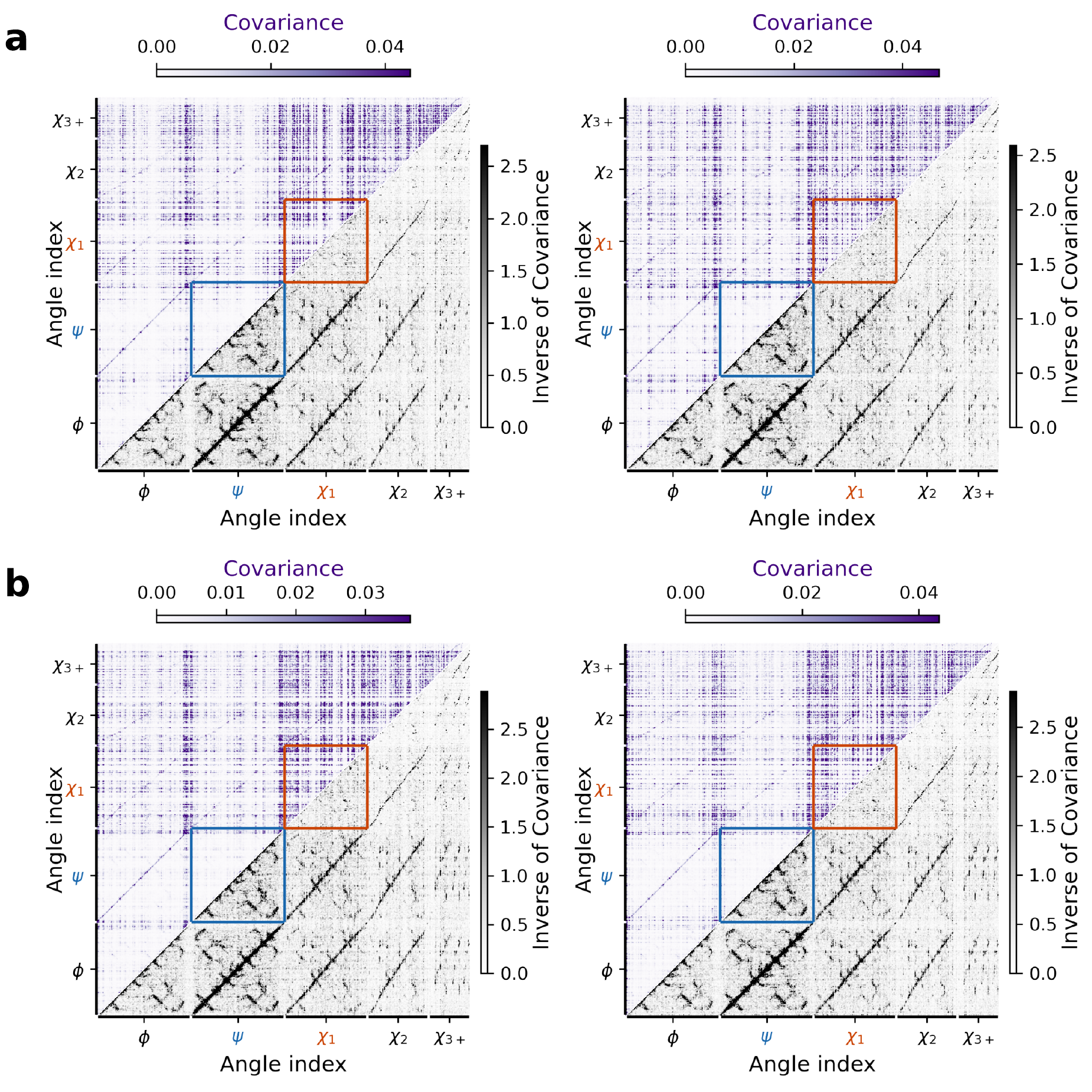}
    \caption{%
    \textbf{The covariance matrix and its inverse show different patterns for Siglec-8.} We show that the inverse covariance matrix resembles the contact map for representative examples in the apo and holo states. 
    Two replicates of Siglec-8 
    \textbf{a}, apo (without ligand, PDBID 2N7A) and 
    \textbf{b}, holo (bound to 6'S-sLe\textsuperscript{x}, PDBID 2N7B)
    See Supplementary Figure \ref{suppfig:CvHex_FimH} for description.
    }
    \label{suppfig:CvHex_Sig8}
\end{suppfigure}

\begin{suppfigure}
    \centering
    \includegraphics[width=\linewidth]{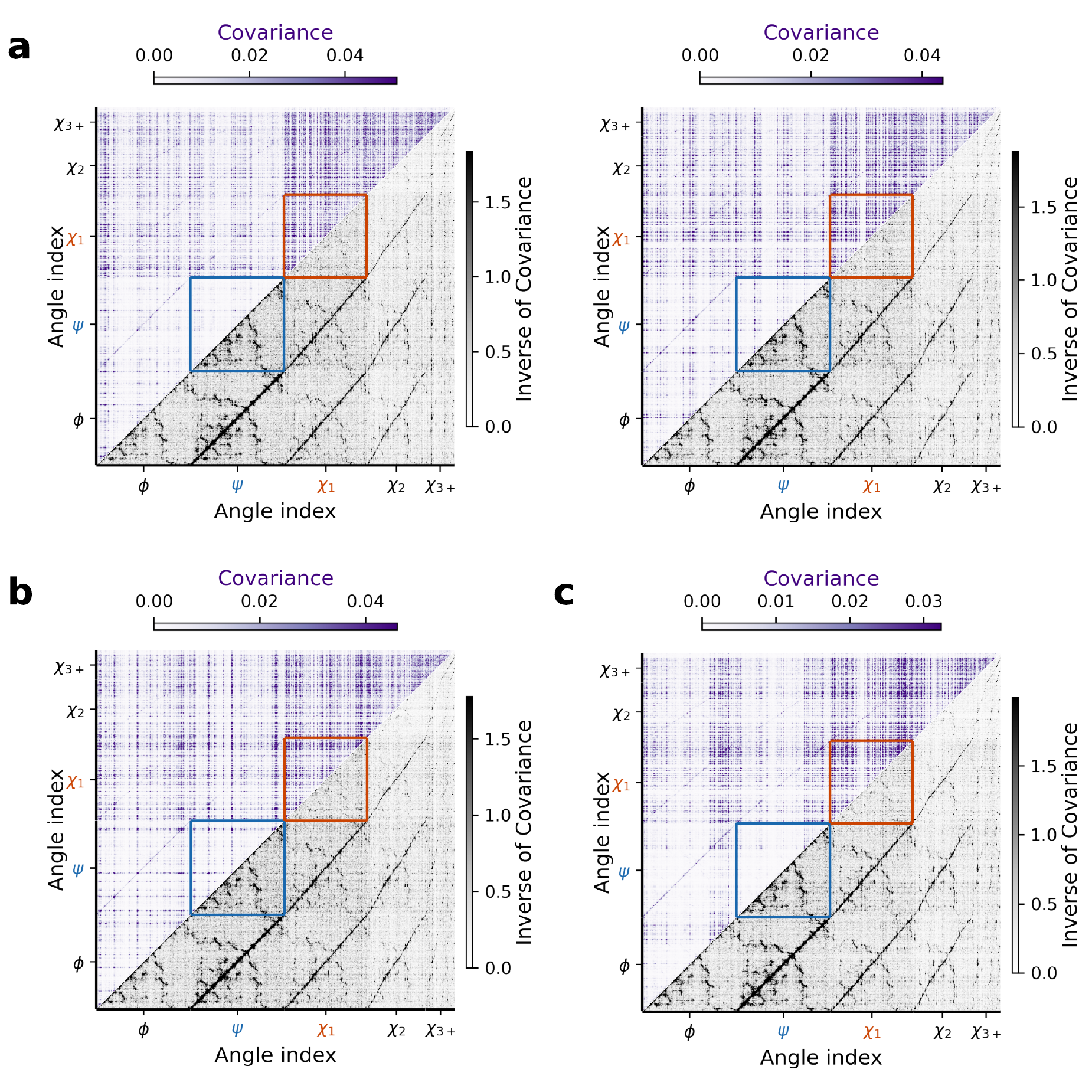}
    \caption{%
    \textbf{The covariance matrix and its inverse show different patterns for SARS-CoV-2.} We show that the inverse covariance matrix resembles the contact map for representative examples of RBD-SD1 domains in the 
    \textbf{a}, up, 
    \textbf{b}, down, 
    \textbf{c}, and off states.
    RBD-SD1 domains in the up and down states are from chains A and B of PDBID 6VSB. The off state is chain A of PDBID 6VXX.
    See Supplementary Figure \ref{suppfig:CvHex_FimH} for description.
    }
    \label{suppfig:CvHex_coronavirus}
\end{suppfigure}

\begin{suppfigure}[ht]
    \centering
    \includegraphics[width=0.9\linewidth]{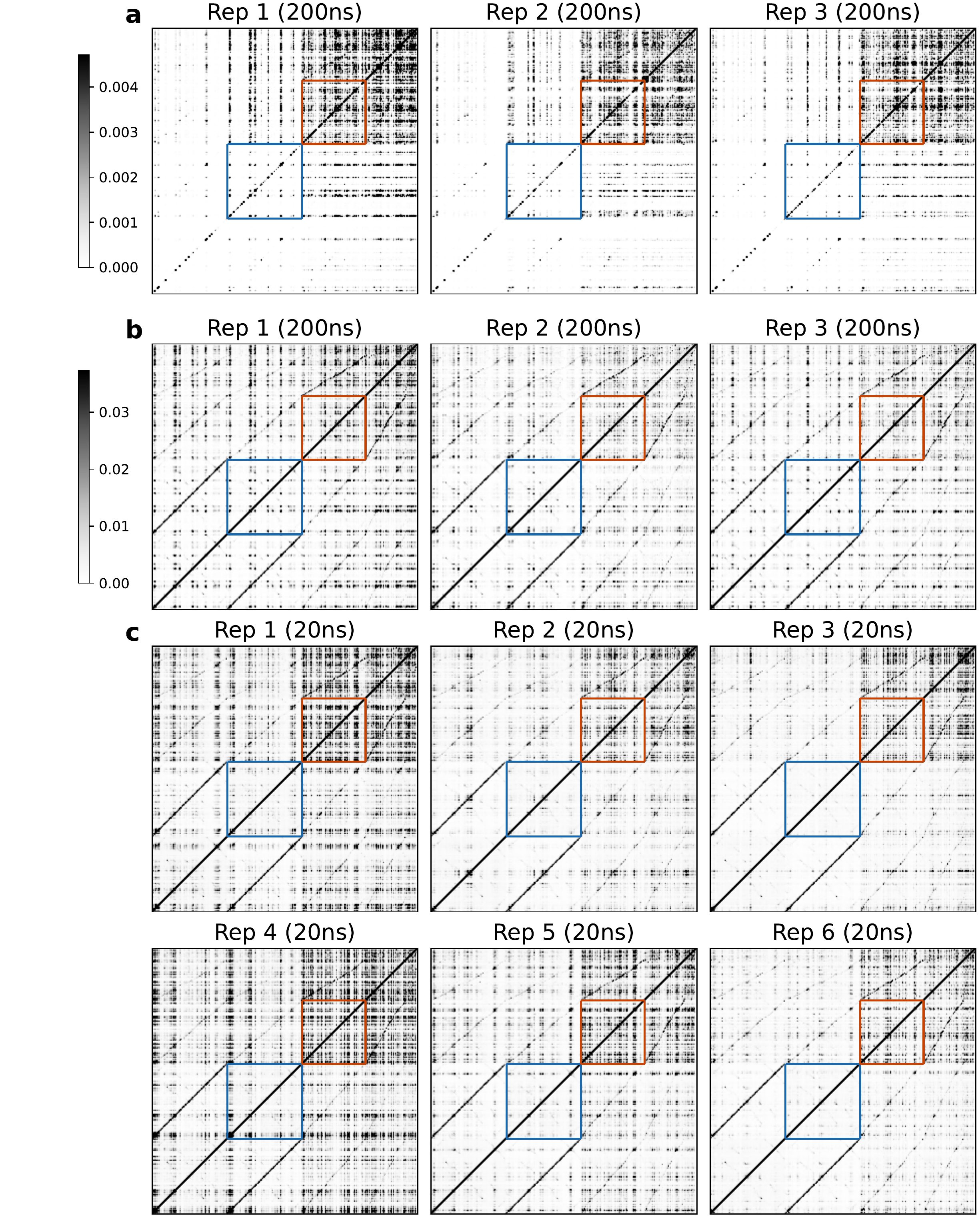}
    \caption{%
    \textbf{Mutual information matrix shows sidechain-sidechain interactions are stronger than backbone-backbone interactions.} We show data for wild-type FimH\textsubscript{L}. Starting with three replicate simulations of 200ns each, we show the mutual information calculated from 
    \textbf{a}, transition state analysis \cite{DuBay2011Long-rangeAlone} with a core of 90 degrees and 
    \textbf{b}, from a histogram based approach. For each pair of angles, we define two sets of unequally spaced bins at the deciles for each individual angle. These two sets of bins are used to construct a 2 dimensional histogram.
    \textbf{c}, We then show six replicate simulations of 20ns. The first three are truncated versions of the longer simulations.
    }
    \label{suppfig:MI_FimH}
\end{suppfigure}

\subsection*{Thresholding for visualization}
While we do not use thresholds for comparing inferred interactions, we do use thresholds to visualize networks as the adjacency matrix and on the protein. In the matrix visualizations, we set the color scale maximum to the 97\textsuperscript{th} percentile. However, to better illustrate the pattern at lower values in the correlation matrix in Figure \ref{fig:similar}A, we also show a lower color scale maximum set to the 95\textsuperscript{th} in Figure S\ref{suppfig:similar_95thpile}. In contrast, for the inverse covariance matrix, showing lower values does not have a profound effect, since the contact map pattern is formed by inferred interactions with high values.

We also show how thresholds affect the Jaccard similarity. In Figure \ref{fig:similar}B, we used a threshold of 97\textsuperscript{th} percentile to create a mask where every edge above the threshold is set to 1, and edges below the threshold are set to 0. Here we show masks for thresholds at the 50, 95, 97, and 99\textsuperscript{th} percentiles, for the covariance, mutual information, and inverse covariance matrices (Figure S\ref{suppfig:thresh_all}). In Figure S\ref{suppfig:similar_jaccard}, we show that for strong interactions, the inverse covariance matrix has high similarity across replicates, while the other matrices show decreasing similarities. While other matrices have higher similarity around the 50\textsuperscript{th} percentile, the masks in Figure S\ref{suppfig:thresh_all} show the extremely large number of edges included at these low thresholds. 

The banded pattern indicates edges connecting one dihedral angle to many others, and an excessive number of these edges produces hairball networks. Thus, for the covariance, correlation, and mutual information matrices, there is a tradeoff where low thresholds with higher similarity produce hairball networks. In contrast, for the inverse covariance network, the high thresholds that produce higher similarity select for stronger interactions, producing physically interpretable networks that resemble the contact map. 

\begin{suppfigure}[]
    \centering
    \includegraphics[width=\linewidth]{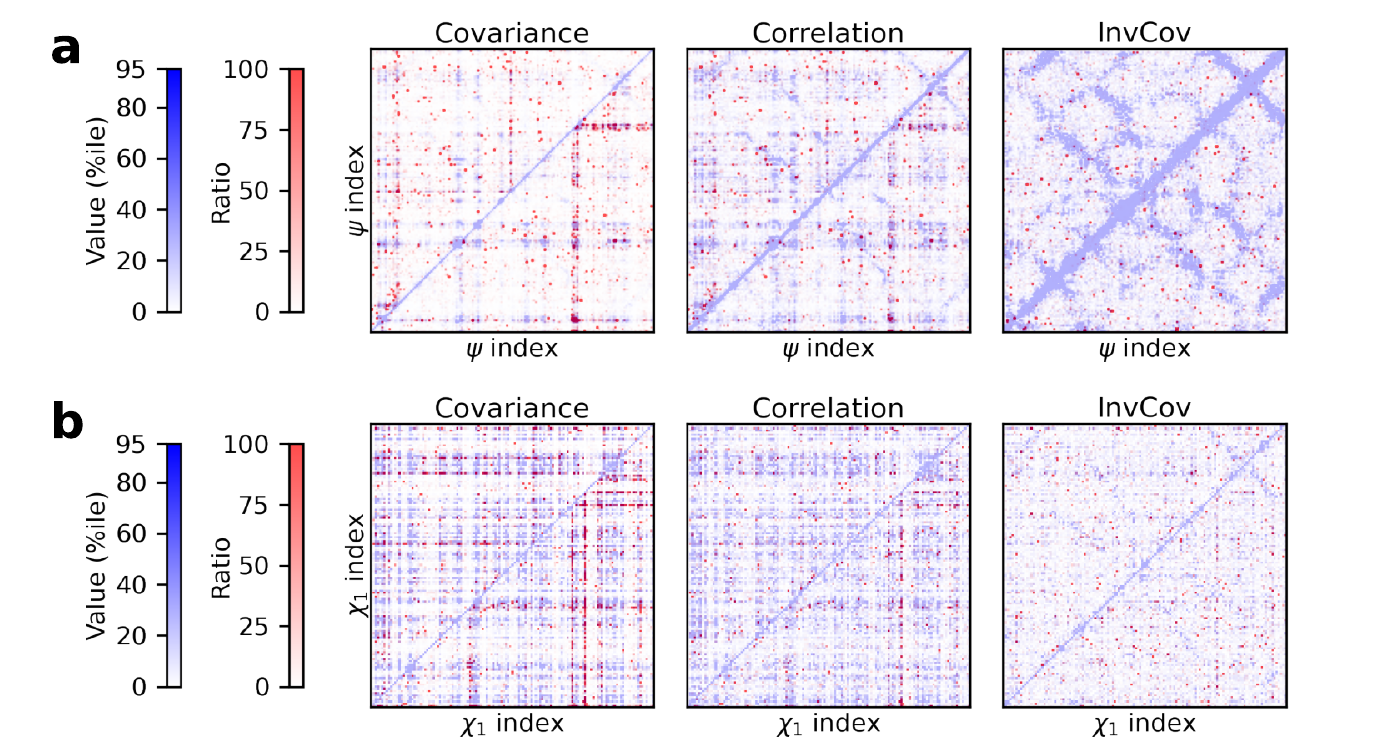}
    \caption{%
    \textbf{Weaker patterns in the backbone correlation and sidechain inverse covariance matrices.}
    \textbf{a}, We show the covariance, correlation, and inverse covariance data from Figure \ref{fig:similar} with the colormap set to the 95\textsuperscript{th} percentile to show weaker interactions. Beneath the banded pattern in the correlation matrix, there is a weaker contact map pattern. 
    \textbf{b}, Corresponding data for $\chi_1-\chi_1$ interactions shows an extremely faint pattern that may resemble a contact in the inverse covariance matrix. 
    }
    \label{suppfig:similar_95thpile}
\end{suppfigure}

\begin{suppfigure}[]
    \centering
    \includegraphics[width=\linewidth]{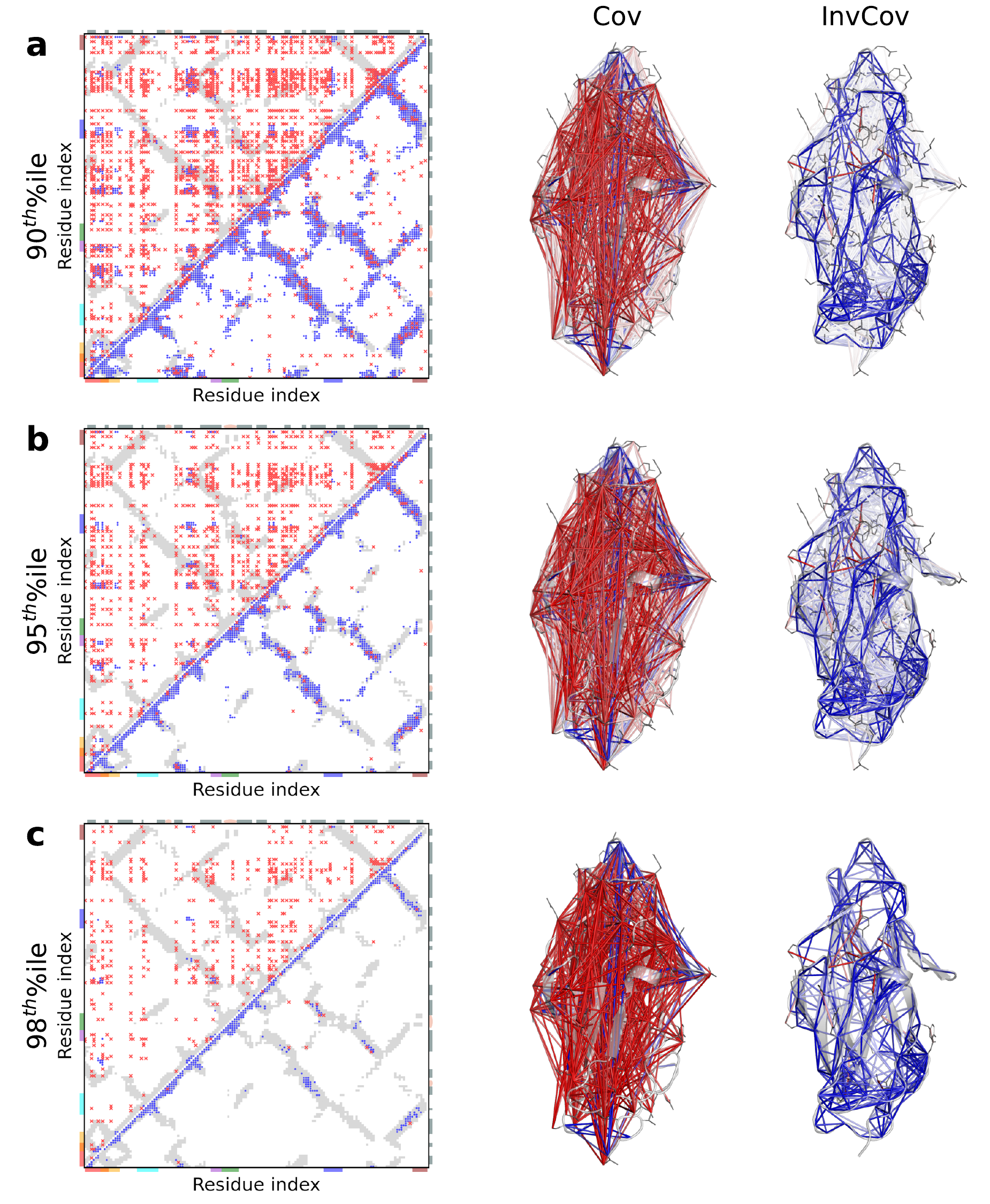}
    \caption{%
    \textbf{The inverse covariance matrix has a ``contact map''-like network at multiple thresholds.} 
    We show the covariance (top left) and inverse covariance (bottom right) at the : 
    \textbf{a}, 90\textsuperscript{th},
    \textbf{b}, 95\textsuperscript{th}, and
    \textbf{c}, 98\textsuperscript{th} percentiles
    to complement the 97\textsuperscript{th} percentile threshold in Figure \ref{fig:net}. 
    We use the same representation scheme as in Figure \ref{fig:net}.
    }
    \label{suppfig:net_thresh}
\end{suppfigure}

\begin{suppfigure}[ht]
    \centering
    \includegraphics[width=\linewidth]{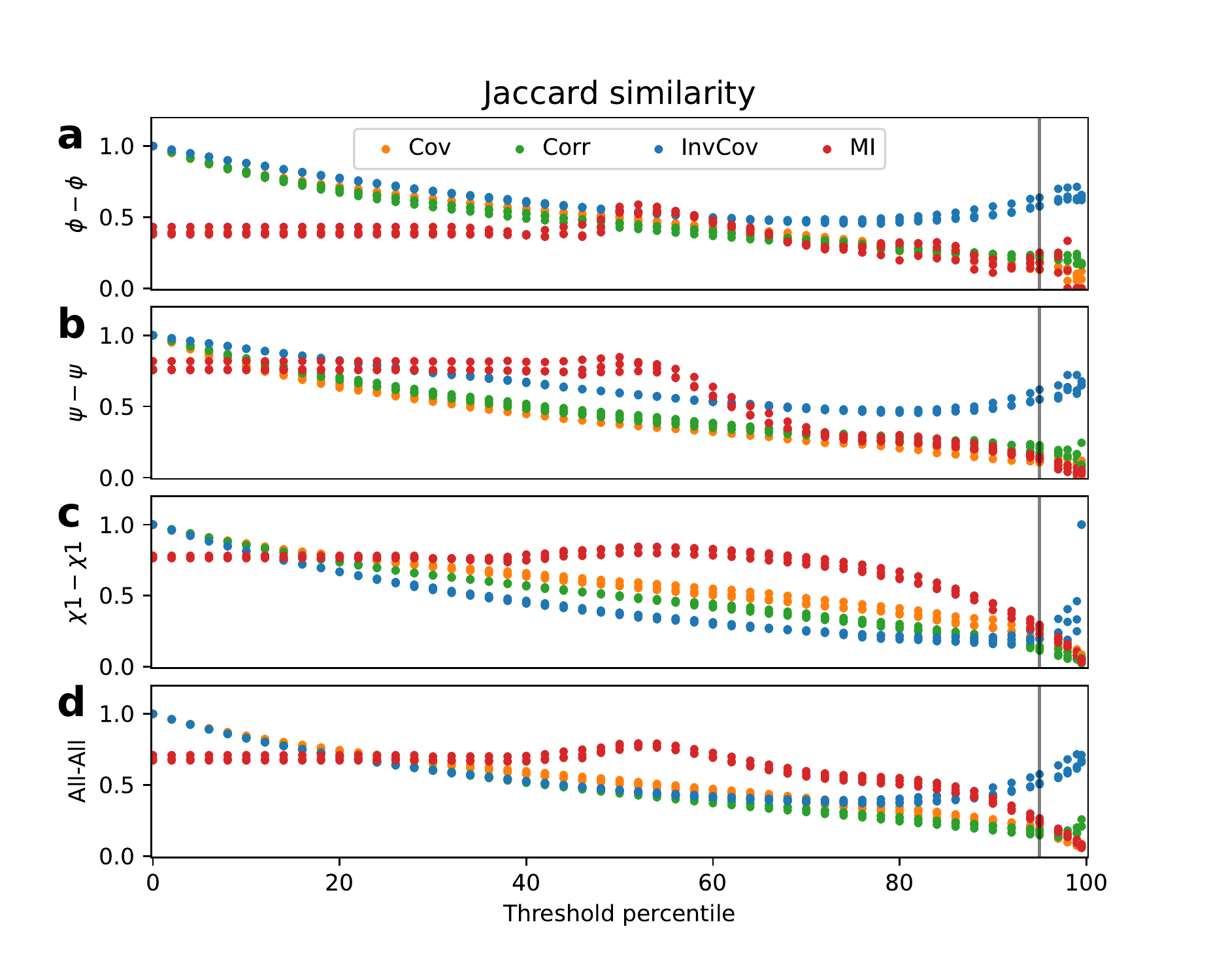}
    \caption{%
    \textbf{Similarity for covariance, correlation, inverse covariance, and mutual information.} We calculate the Jaccard similarity for the networks defined by these four methods at different thresholds. We do this for the backbone-backbone and sidechain-sidechain interactions, and also for the entire protein. The vertical line indicates the 97\textsuperscript{th} percentile used in Figure \ref{fig:similar}B.
    See Supplementary Figure \ref{suppfig:thresh_all} for adjacency matrix representations at representative thresholds. 
    }
    \label{suppfig:similar_jaccard}
\end{suppfigure}

\begin{suppfigure}[ht]
    \centering
    \includegraphics[width=\linewidth]{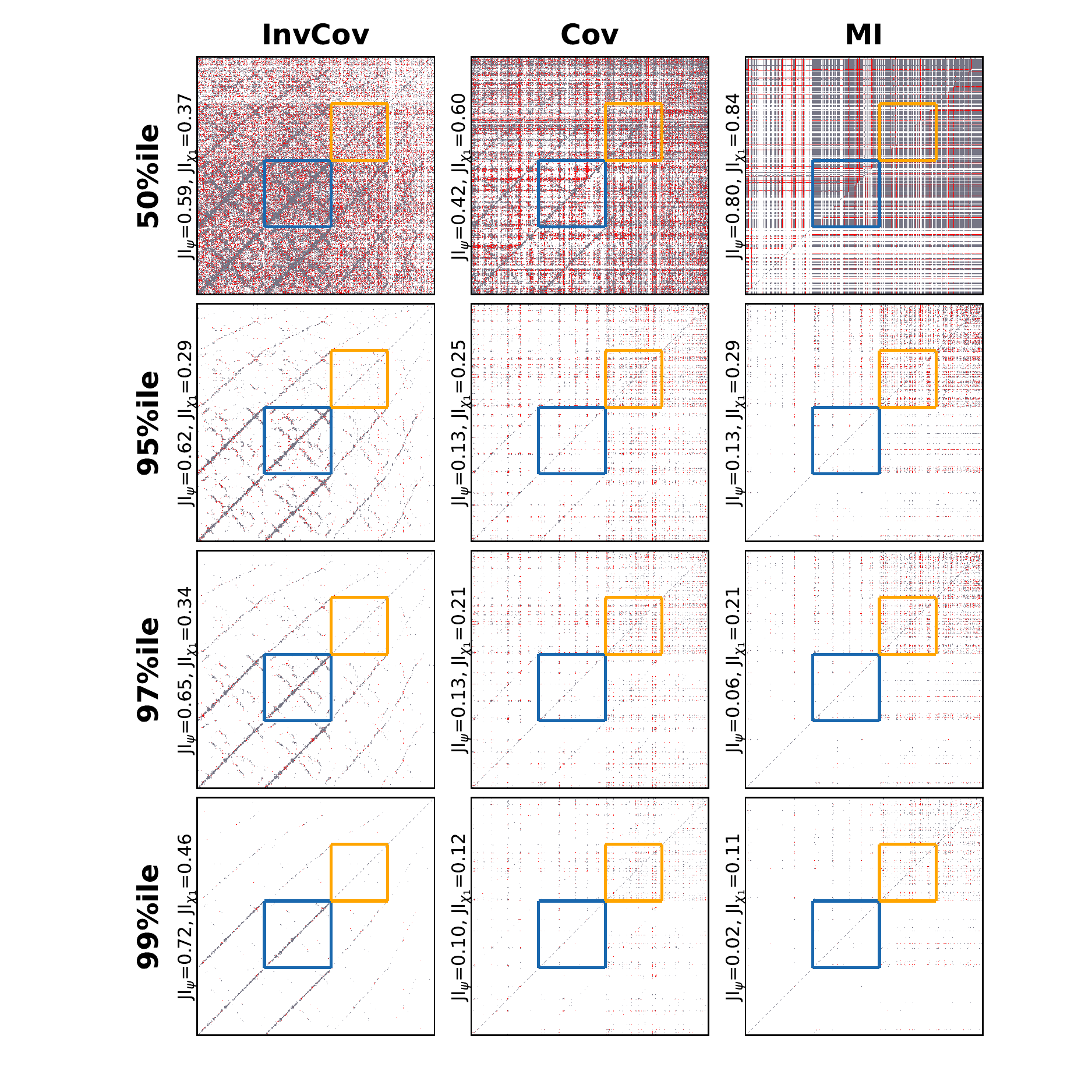}
    \caption{%
    \textbf{Representative thresholds for defining unweighted networks used to calculate the Jaccard Index (JI) for the inverse covariance, covariance, and mutual information matrices.}
    We directly compare two replicates in the top left and bottom right triangles. Edges above the threshold in both replicates are shown in grey, while those only present in one replicate are shown in red. For each replicate pair, we list JI for $\psi-\psi$ (blue box) and $\chi_1-\chi_1$ (yellow box) submatrices.
    }
    \label{suppfig:thresh_all}
\end{suppfigure}

\begin{suppfigure}[]
    \centering
    \includegraphics[width=\linewidth]{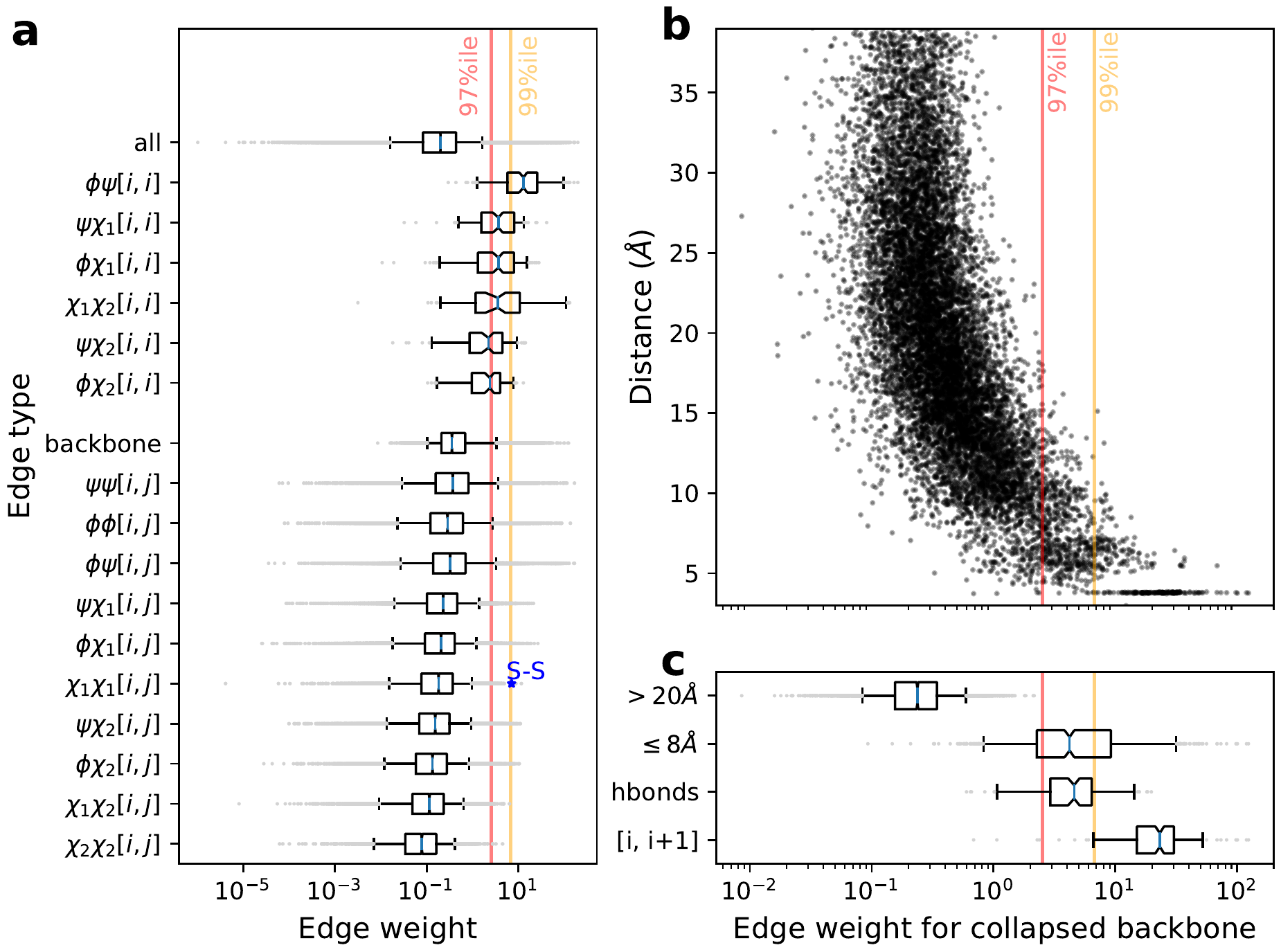}
    \caption{%
    \textbf{Hierarchy of interaction strengths suggest a multilayer network.} 
    \textbf{a}, For qualitatively different interaction types, we show the distribution of interaction strengths. We compare interactions within the same residue for different dihedral angles [i, i], as well as interactions between different residues for the same dihedral angles [i, j]. The backbone label indicates the collapse of backbone layers, as calculated from the mean interaction strength for $\phi-\phi$, $\psi-\psi$, $\phi-\psi$, and $\psi-\phi$ interactions. The box-and-whisker plots mark the 5, 25, 50, 75, and 95\textsuperscript{th} percentiles, with the median in blue, and outliers in grey. The Cys3-Cys-44 disulfide bond is shown with a star in dark blue. 
    \textbf{b}, For the collapsed backbone interactions, we show the relationship between edge strength and C$\alpha$-C$\alpha$ distance. Stronger interactions are associated with smaller distance. 
    \textbf{c}, We show distributions of interaction strengths for residue pairs with C$\alpha$-C$\alpha$ distances that are far apart ($\geq 20$\AA) and close together ($\leq 8$\AA), residue pairs with backbone hydrogen bonds, and neighbors on the primary sequence (i, i+1).
    }
    \label{suppfig:multi_edge}
\end{suppfigure}

\begin{suppfigure}[ht]
    \centering
    \includegraphics[width=\linewidth]{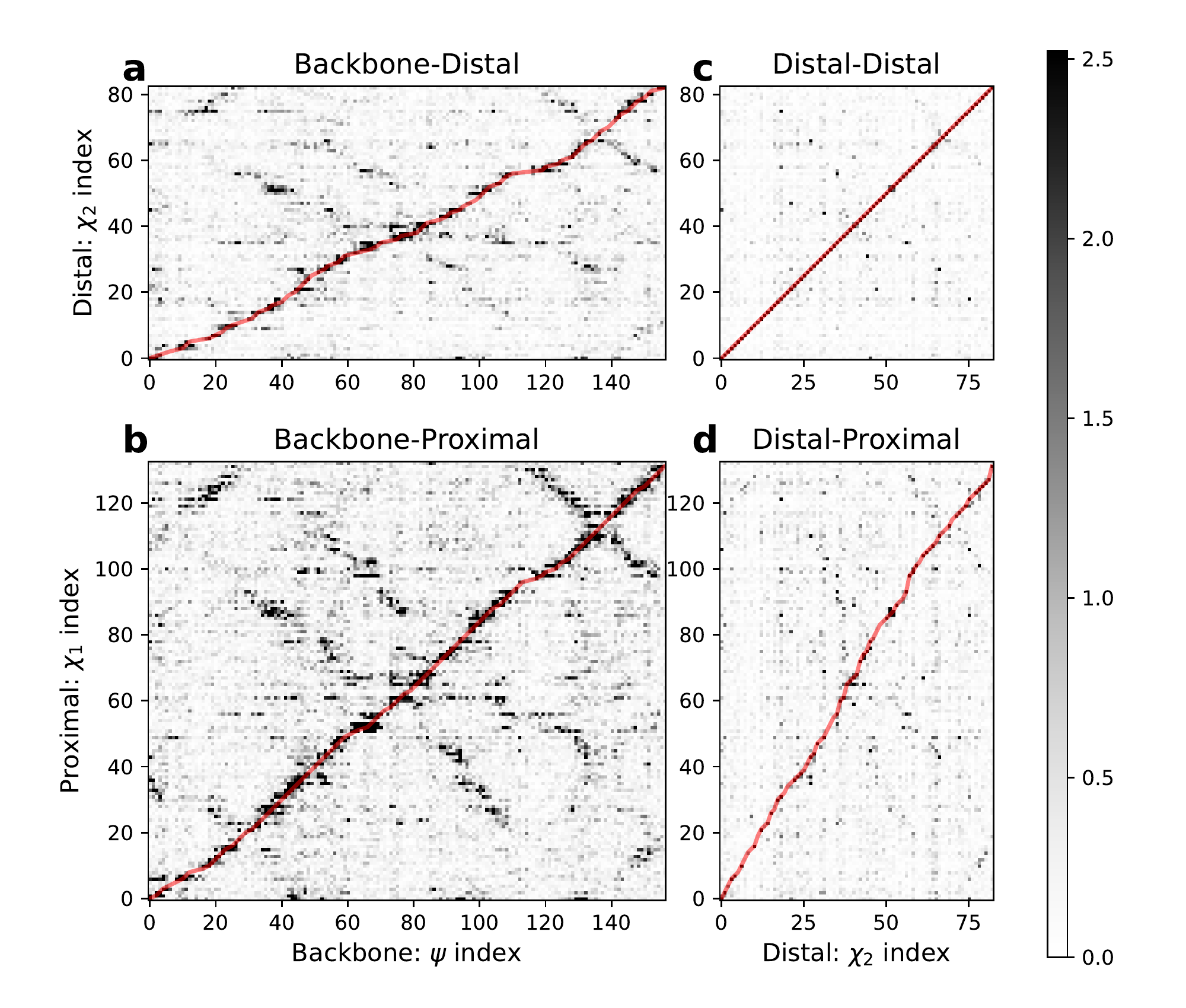}
    \caption{%
    \textbf{Sidechain interactions become weaker further away from the backbone.} Grey scale indicates absolute value of the interaction strength for the inverse covariance for one replicate of wild type FimH\textsubscript{L}.
    \textbf{a,}
    Backbone dihedral ($\psi$) interactions with a distal sidechain dihedral ($\chi_2$) is weaker than
    \textbf{b, }
    interactions with a more proximal one ($\chi_1$). $\psi-\chi_1$ interactions still retains the contact map pattern.
    \textbf{c, }
    While interactions between distal sidechains $\chi_2-\chi_2$ do not have a clear pattern,
    \textbf{d, }
    yet, there is a still a slight pattern between proximal and distal sidechains ($\chi_1-\chi_2$).
    }
    \label{suppfig:multi_pattern}
\end{suppfigure}

\begin{suppfigure}[ht]
    \centering
    \includegraphics[width=\linewidth]{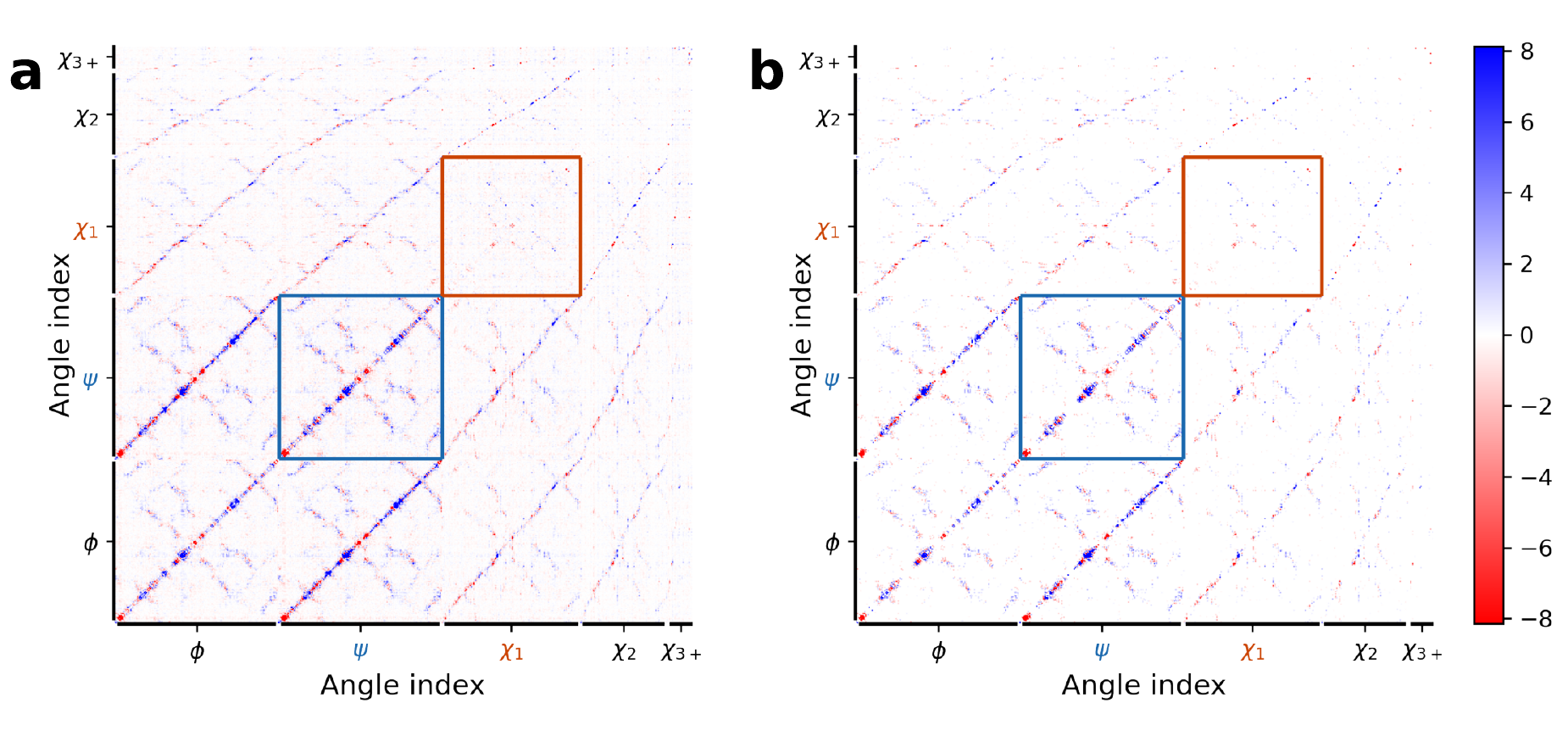}
    \caption{%
    \textbf{Comparison of inferred networks for wild type and mutant FimH\textsubscript{L} using all inferred edges.} 
    For wild type - mutant, we show edges stronger in the wild type in blue, and those for the mutant in red (colorbar). We show two versions: 
    \textbf{a}, without thresholding and 
    \textbf{b}, requiring differences to be larger than twice the standard deviation of technical replicates within each group. We do not apply any filters based on distance, secondary structure, or other structural information.
    }
    \label{suppfig:WTvsMUT_all}
\end{suppfigure}


\begin{suppfigure}[]
    \centering
    \includegraphics[width=\linewidth]{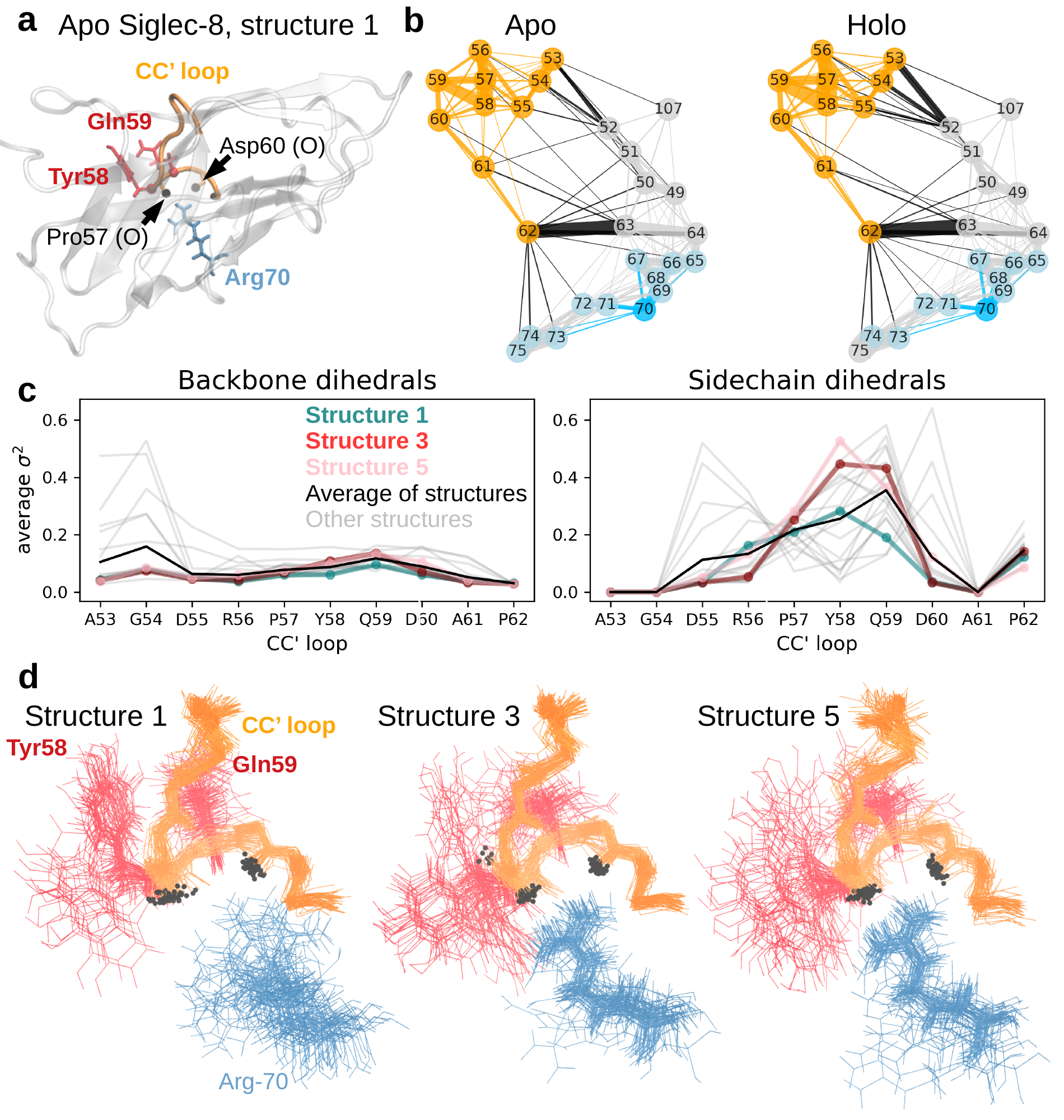}
    \caption{%
    \textbf{Steric hindrance as a potential mechanism for CC' loop stability in apo Siglec-8.} 
    \textbf{a}, Illustration of apo Siglec-8 highlighting the CC' loop (orange) and the carbonyl O atoms of Pro57 and Asp60 (black), which are hypothesized to be hydrogen bond acceptors for Arg70 (blue). We show residues Tyr58 and Gln59 (red) at the tip of the loop, between the hydrogen bond acceptors. 
    \textbf{b}, Inferred networks for apo and holo Siglec-8 showed interactions within the CC' loop, between the edges of the CC' loop and other residues, but not interactions with Arg70. We show the average network for all 20 structures from the ensemble. 
    \textbf{c}, Average variance of dihedral angles for the backbone and sidechains of the CC' loop for apo Siglec-8. In 50ns MD simulations of structures 3 and 5, Arg70 formed hydrogen bonds with Pro57 and Asp60, while it did not in structure 1. Tyr58 and Gln59 in the CC' loop have larger dihedral fluctuations for structures 3 and 5 (red and pink) than for 1 (blue).
    \textbf{d}, Visualization of the dynamics as overlaid 1ns snapshots show larger fluctuations for Arg70 in structure 1, while structures 3 and 5 have larger fluctuations for Tyr58 and Gln59.
    }
    \label{suppfig:sig8CCloophbond}
\end{suppfigure}

\begin{suppfigure}[ht]
    \centering
    \includegraphics[width=0.95\linewidth]{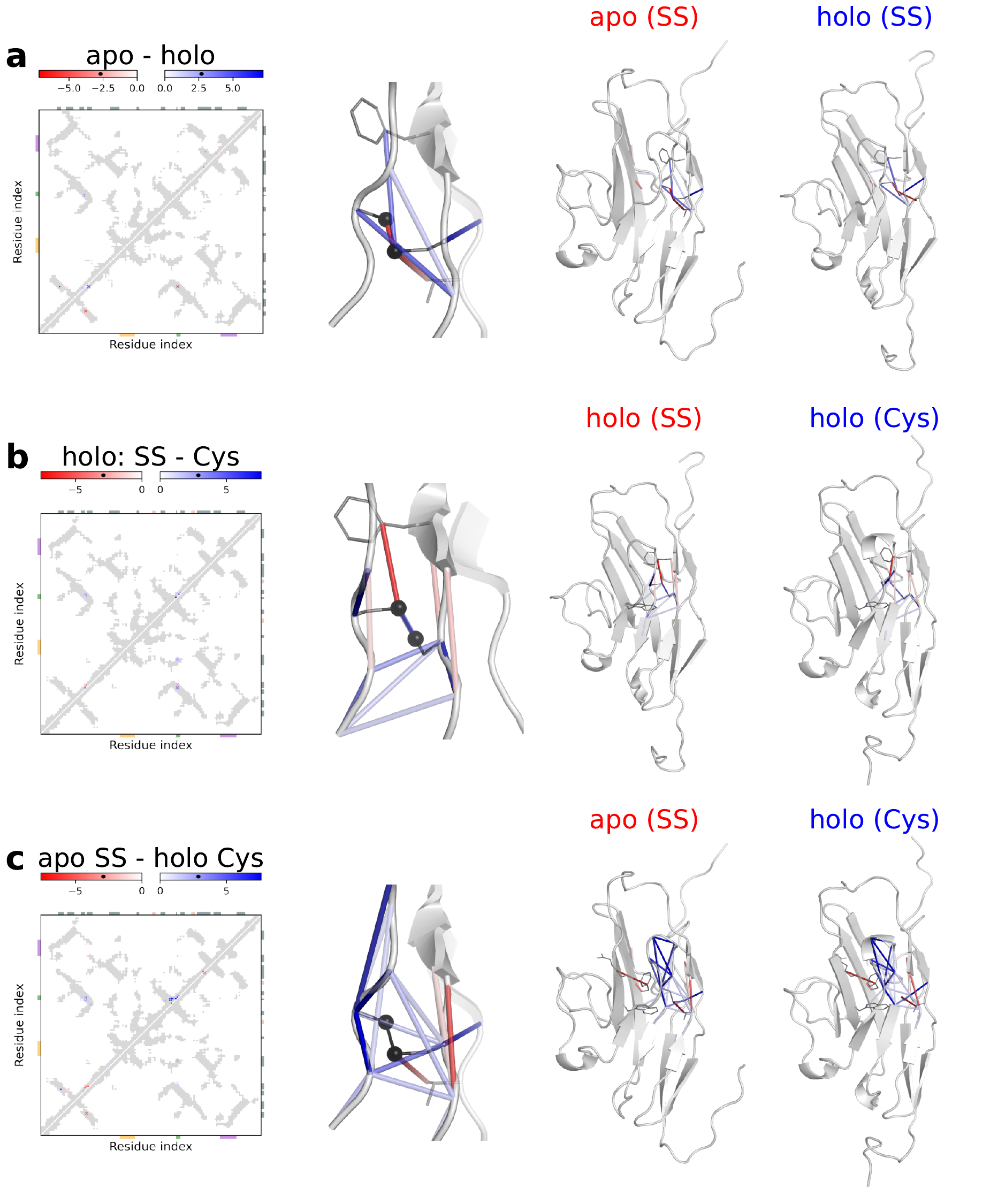}
    \caption{\textbf{Impact of ligand-binding and reducing the Cys31-Cys91 disulfide bond on Siglec-8.}
    Difference in inferred network interactions for 
    \textbf{a}, apo vs holo states, and
    \textbf{b}, the holo state with and without the disulfide bond. Due to the difference in inferred interaction strength at the disulfide bond between apo and holo Siglec-8, we also compared 
    \textbf{c}, the apo state with the disulfide bond intact vs the holo state without the disulfide bond. 
    We use the same representation scheme as in Figure \ref{fig:fimh}. Differences in the backbone-backbone interactions are shown in the top left in dots; sidechain-sidechain interactions in the bottom right as crosses. On the adjacency matrix, we show all differences between groups larger than 2$\sigma$, but we only draw differences larger than the 97\textsuperscript{th} threshold for interaction strength on the protein structure. 
    We show the secondary structure on the top and right borders. We show landmarks on the bottom and left borders: CC' loop (orange), GG' loop (purple), and the evolutionarily conserved Asp90-Cys91-Ser92 motif (green). 
    }
    \label{suppfig:siglecall}
\end{suppfigure}


\begin{suppfigure}[]
    \centering
    \includegraphics[width=\linewidth]{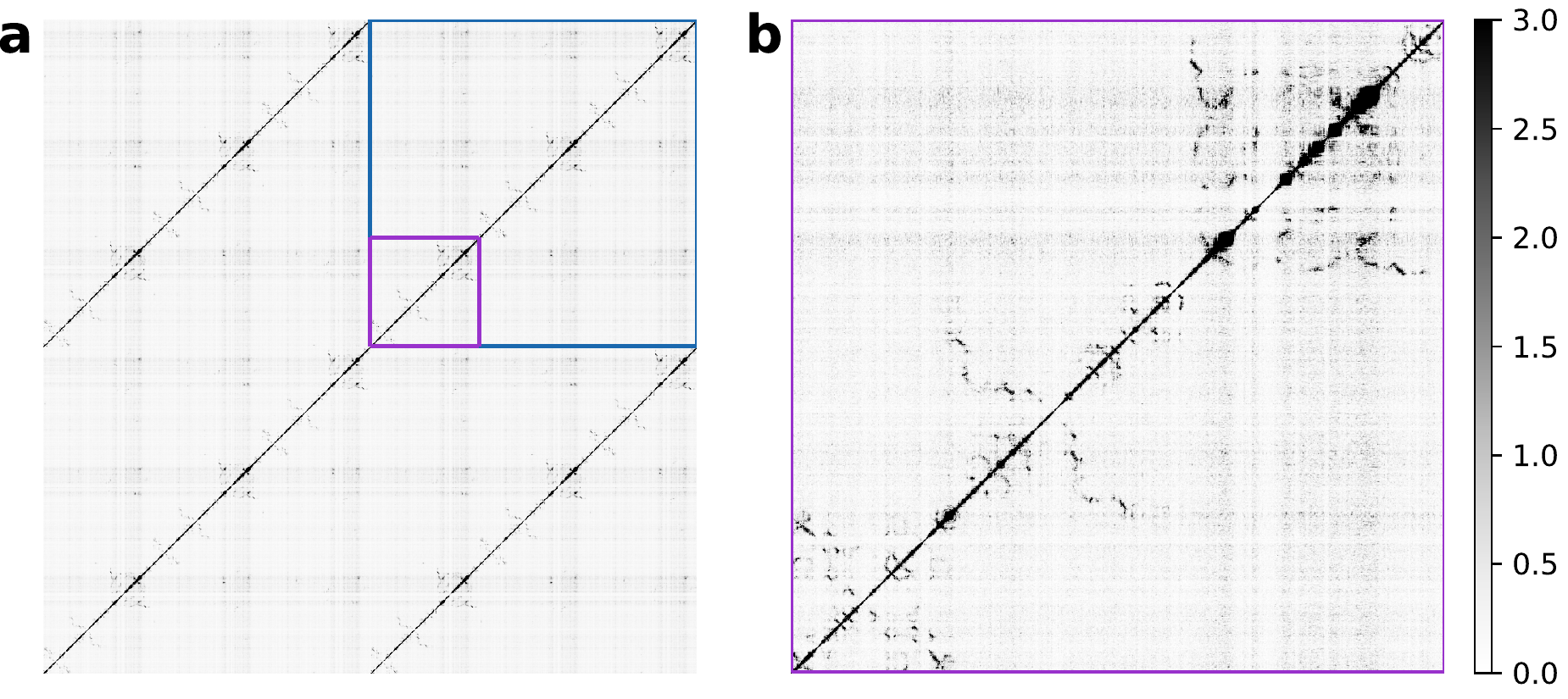}
    \caption{\textbf{Network inference for the trimeric spike protein.}
    \textbf{a}, Backbone-backbone interactions for the trimer, with $\psi$-$\psi$ interactions highlighted by the blue box. The first protomer is shown with a purple box, to highlight the pattern that repeats three times within the blue box. The other backbone-backbone interactions also show a pattern that repeats three times. 
    \textbf{b}, 
    Zoomed-in view of the first protomer's $\psi$-$\psi$ interactions shows a contact-map pattern. 
    }
    \label{suppfig:spike_H}
\end{suppfigure}

\begin{suppfigure}[]
    \centering
    \includegraphics[width=\linewidth]{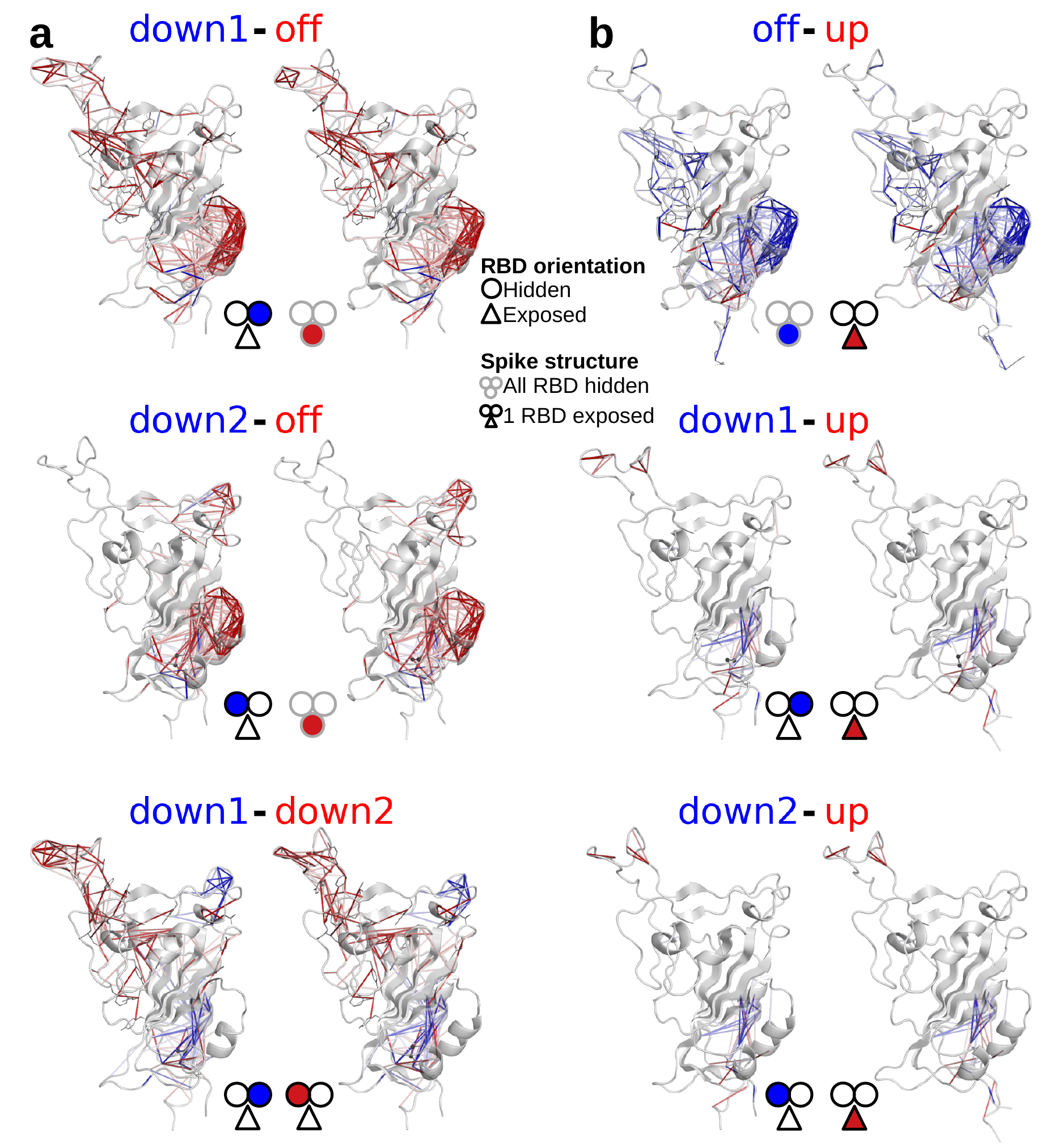}
    \caption{\textbf{Impact of up, down, or off state for the SARS-CoV2-spike protein RBD-SD1 domains.} In the up state (PDBID 6VSB, chain A), the RBD is accessible to bind human ACE2 for viral attachment. The neighboring protomers (chains B and C) have hidden RBDs. The off state (PDBID 6VXX, chain A) comes from a homotrimer with 3-fold rotational symmetry, where all RBDs are hidden. 
    Difference in inferred network interactions 
    \textbf{a}, among the down and off states, 
    \textbf{b}, and with the up state.
    We use the same representation scheme as in Figure \ref{fig:fimh} and Figure S\ref{suppfig:siglecall}. 
    }
    \label{suppfig:RBDSD1all}
\end{suppfigure}

\end{document}